\documentclass[11pt,a4paper,notitlepage]{article}

\usepackage{a4wide}

\usepackage{enumitem}
\usepackage{epsfig}
\usepackage[cmex10]{amsmath}
\usepackage{amsfonts}
\usepackage{mathrsfs}
\usepackage{alltt}
\usepackage{multicol}
\usepackage{csvsimple}
\usepackage[table]{xcolor}

\usepackage{authblk}

\usepackage[english]{babel}

\title{Automatic Compiler-Based Data Structure Generation}

\author{Kristian~F.~D.~Rietveld\thanks{E-mail: krietvel@liacs.nl}~}
\author{Harry~A.~G.~Wijshoff\thanks{E-mail: harryw@liacs.nl}}
\affil{LIACS, Leiden University, The Netherlands}

\date{}

\begin{document}

\maketitle
\begin{abstract}
Optimizing compilers are mainly equipped to optimize control flow.  The
optimization of data structures is left to the programmer and it is the
programmer's responsibility to design the data structures to suit the target
hardware. Very specific data structures are required to exploit certain
hardware features, such as cache line size, address alignment, vector width,
and memory hierarchy specifics. Because optimizing compilers do not target
data structures, these features are explicitly encoded in program
specifications.  This leads to convoluted programs that obscure the
essence of the computation from the compiler, in turn causing compiler
analysis techniques to break down and hampering compiler optimizations from
being applied.

To solve this problem, we propose to move towards the specification of
programs without explicitly specifying the data structure. The compiler will
automatically generate actual data structures and executable code starting
from this specification. In this paper, we introduce a compiler-based
framework to support this automatic generation of data structures, allowing
the compiler to go beyond the optimization of solely control flow and also
target the way data is organized and accessed. As a case study of the
effectiveness of this framework, we present a detailed description and
experimental results of the application of the proposed techniques to
automatically generate data structures for sparse matrix computations. We
show that this way sparse data structures can be generated that were up till
now only specified by hand and that automatically instantiated routines and
corresponding data structures can be found that outperform implementations
of three existing sparse algebra libraries.
\end{abstract}

\noindent
{\bf Keywords:} Data Structure Generation, Code Generation,
Program Transformation, Optimizing Compilers, Sparse Matrices

%

\section{Introduction}
Optimizing compilers have traditionally been used as tools to optimize
control flow. For instance, transformations such as common subexpression and
dead code elimination remove superfluous instructions from the control
flow. Loop transformations modify the flow of control within
loops to better utilize the cache by re-using cached data~\cite{padua-1986}.
Subsequently, in the compiler's code generation stage, the operations that
make up the control flow are translated to machine code.  During this
translation, (combinations of) machine code instructions can be specifically
selected to improve performance.
Common to all of these approaches is the fact that the compiler is equipped
with tools to optimize control flow.  The compiler does not touch the data
structure as has been specified in the program specification at all.
Traditional compiler optimizations target the instructions that manipulate
the data, but do not modify how the data is organized.

In other words, because optimizing compilers are mainly equipped with tools
to optimize control flow, the optimization of data structures is left to the
programmer.  It is the programmer's responsibility to design the data
structures to suit the hardware, taking into account temporal and spatial
locality, cache line size, size of the cache, memory address alignment,
vector width, etc.
As a consequence of explicitly exploiting these features within (very)
specific data structures in program specifications, the essence of the
computation is obscured from the compiler. The use of pointer-linked data
structures and indirection arrays cause compiler analysis techniques to
break down and hampers compiler optimizations from being applied. So, the
lack of capability of optimizing compilers to target data structures causes
programmers to write code that is harder, or impossible, to analyze by these
compilers.

To solve this problem, programs must be specified in a way such that details
of the computation are not obscured from the compiler. To this end, we
propose to move towards the specification of programs without explicitly
specifying the data structure. Starting from this specification, the
compiler will automatically generate an actual data structure and executable
code that operates on this data structure. Through the use of
transformations, the data structure and code that are generated can be
modified.

In this paper, we introduce a compiler-based framework to support this
automatic generation of data structures. This framework allows the compiler
to go beyond the optimization of solely control flow also targeting the way
data is organized and accessed. The cornerstone of this framework is a
different scheme to specify programs. Within this scheme, specification of
fixed iteration order and the fixed specification of data structures is
avoided.  By doing so, the compiler is given much more freedom to perform
optimizations, as the compiler is no longer obstructed by fixed
specifications of codes and data structures that it cannot violate.

Instead of the traditional specification of data structures, within this
framework tuples are used as elementary data structure and programs are
written as manipulations of sets of tuples. In fact, traditional data
structures are ``disassembled'' into tuples. Tuples have been selected as
elementary data structure because tuples are the most fundamental structures
to relate data to one another. So, data that is related to one another is
grouped in tuples. No such groups are created for unrelated data,
suppressing the specification of false, or unintended, relationships.
Additionally, consider that computer memory can be represented in terms of
tuples that pair an address and value. From this follows that all possible
data structures can be expressed using tuples and also that all possible
data structures can be generated from a tuple-based specification.

Within the framework that is proposed, transformations are defined that are
applied on the programs that manipulate sets of tuples. These
transformations affect both the order in which data is visited by the actual
computation (control flow) and the order in which data is stored (data
storage). After the application of these transformations a new stage is
entered in which a data structure is generated automatically according to
the current state of the computational loop. Hence, a computation-driven
reassembly of the data structure is carried out leading to a data structure
to be composed from tuple-based storage. Different chains of transformations
lead to different computational loops that lead to different data structures
to be generated. So, within this scheme the computation and data structure
are optimized hand in hand.

As a case study, we present a detailed description and experimental results
of the application of the proposed techniques to automatically generated
data structures for sparse matrix computations. Since sparse matrix
computations are an important class of compute intensive codes, many
techniques have been developed to optimize these computations. A large
majority of these techniques concerns the creation of smart data structures
for storing the sparse matrix data. This approach exposes two major
drawbacks: firstly, these data structures are to be specified by the
programmer, as the compiler cannot optimize data organization and secondly,
by wrapping the sparse matrix data in a specific data structure, the
compiler optimization process is obstructed. We show that by using the
automatic data structure generation approach, sparse data structures can be
generated automatically that were up till now only specified by hand.
The transformations described in this paper lead to a large search space of
possible code variants: essentially 25 different data structures are being
generated. This exemplifies the strength of our approach when compared to
sparse algebra libraries, that on average implement 4, or less, pre-defined
sparse data formats. In an experimental evaluation of the generated codes
and data structures, we show that in this search space automatically
instantiated routines and corresponding data structures can be found that
outperform implementations of three existing sparse algebra libraries.
Moreover, we show that none of these three libraries performs consistently
better, which means that by relying on a single library implementation
performance is never optimal, further reinforcing the validity of our
approach.

This paper is organized as follows. In Section~\ref{sec:rationale} we give a
rationale for the approach described in this paper.
Section~\ref{sec:forelem-intro} presents a brief introduction of the
\emph{forelem} framework.
Section~\ref{sec:transformations} describes transformations within the
\emph{forelem} framework that enable automatic data structure generation:
using the Orthogonalization transformation an explicit order on iteration
can be imposed, Materialization will materialize the loop to a single
particular execution order, subsequently a number of transformations are
described that can be applied on \emph{forelem} loops in the materialized
form.
Section~\ref{sec:ds-gen} demonstrates how different data storage formats can
be generated by integrating materialization with other transformations
defined in the \emph{forelem} framework.
Section~\ref{sec:case-study} evaluates the framework introduced in this
paper in the context of sparse BLAS kernels. It is shown how
different sparse BLAS kernels can be expressed in the \emph{forelem}
framework and how the transformations proposed in this paper will lead to
the automatic generation of different sparse data structures.
Section~\ref{sec:related-work} discusses related work and
Section~\ref{sec:conclusions} presents our conclusions and plans for future
work.

\section{Rationale}
\label{sec:rationale}
The lack of compiler transformations that affect data organization forces
programmers to implement hand-optimized data structures in the program
specification. The main problem with this is that this commonly leads to
complicated data structures, which cause compiler analysis techniques to
break down. As an example, consider a graph stored as an array of edges
\verb!(u, v, w)! and a loop to compute the average weight of edges out of a
vertex \verb!X!. This could be written as follows:

\small
\begin{alltt}
sum = 0;
count = 0;
for (int i = 0; i < N; i++)
\{
  if (edges[i].u == X)
  \{
    count++;
    sum += edges[i].w;
  \}
\}
avg = sum / count;
\end{alltt}
\normalsize

\noindent
This loop with static loop bounds and simple direct array accesses is very
well analyzable by optimizing compilers. The main problem with this code is
the selection of the data structure for this problem. For very large graphs,
the array \verb!edges! will be very large but still needs to be iterated in
full. Here we see the inherent problem of contemporary compilers that only
optimize control flow: these compilers can speed up the execution of the
loop that iterates through the full array, but the compiler cannot modify
the data structure such that only data of interest to the problem can be
visited.

To alleviate this problem, the programmer will manually modify the data
structure used in the program. For instance, consider the programmer creates
a linked list containing the out edges for each node. This leads to the
following code, where \verb!edge_list! is an array of linked list head
pointers, subscripted using the vertex number:

\small
\begin{alltt}
sum = 0;
count = 0;
for (List *l = edge_list[X]; l != NULL; l = l->next)
\{
  count++;
  sum += l->w;
\}
avg = sum / count;
\end{alltt}
\normalsize

\noindent
This code will show much better performance. But it comes at a significant
cost: the loop bounds are now dynamic and data accesses are done through
pointers. As a consequence, compiler analysis breaks down and the compiler
is limited in the amount of optimization it can perform. In case the above
loop is directly expressed in \emph{forelem}, then we can automatically
generate at least X different versions, see Figure~\ref{fig:list-variants}.

\begin{figure}
\scriptsize
\begin{minipage}[t]{0.49\linewidth}
\begin{alltt}

\textrm{\textbf{Array iteration}}

\textbf{for} (int i = 0; i < N; i++)
  \textbf{if} (edges[i].u == X)
  \{
    count++;
    sum += edges[i].w;
  \}

\textrm{\textbf{Array iteration, Orthogonalized on \texttt{u}}}

\textbf{for} (int i = 0; i < N[X]; i++)
\{
  count++;
  sum += edges[X][i].w;
\}

\textrm{\textbf{Array iteration with mask}}

\textbf{// Initialization}
mask[] = FALSE;
\textbf{for} (int i = 0; i < N; i++)
  \textbf{if} (edges[i].u == X)
    mask[i] = TRUE;

\textbf{// Iteration}
\textbf{for} (int i = 0; i < N; i++)
  \textbf{if} (mask[i])
  \{
    count++;
    sum += edges[i].w;
  \}

\textrm{\textbf{Array iteration with set}}

\textbf{// Initialization}
\textbf{for} (int i = 0; i < N; i++)
  \textbf{if} (edges[i].u == X)
    set.insert(i)

\textbf{// Iteration}
\textbf{for} (it = set.begin(); it != set.end(); it++)
\{
  count++;
  sum += it->w;
\}
\end{alltt}%
\end{minipage}%
\begin{minipage}[t]{0.49\linewidth}
\begin{alltt}

\textrm{\textbf{Parallelized array iteration with value-based\\orthogonalization}}

\textbf{forall} (int i = 0; i <= M; i++)
  \textbf{for} (int j = 0; j < N[i]; j++)
    \textbf{if} (edges[i][j].u == X)
    \{
      count++;
      sum += it->w;
    \}

\textrm{\textbf{Linked list iteration}}

current = start;
\textbf{while} (current != NULL)
\{
  \textbf{if} (current->u == X)
  \{
    count++;
    sum += current->w;
  \}

  current = current->next;
\}

\textrm{\textbf{Orthogonalized on \texttt{u}, linked list iteration}}

current = start[X];
\textbf{while} (current != NULL)
\{
  count++;
  sum += current->w;

  current = current->next;
\}

\textrm{\textbf{Orthogonalized on \texttt{u}, array iteration}}

current = start;
\textbf{while} (current != NULL)
\{
  \textbf{if} (current->u == X)
    \textbf{for} (int i = 0; i < current->N; i++)
    \{
      count++;
      sum += current.edges[i];
    \}
\}

\end{alltt}
\end{minipage}
\normalsize
\caption{Different versions that can be automatically generated from a
\emph{forelem} representation of the loop computing average weight of the
out edges of a vertex \texttt{X}}%
\label{fig:list-variants}
\end{figure}

\noindent
Note that in the above example, the traversal has been coded to be performed
in a specific order, while in this case it is safe to visit the elements in
any order. A program specification should not be unnecessarily restricting
the control flow to a specific execution order, as this again hampers the
compiler from generating more efficient code.

\subsection{Data structure-less Programming}
In order to facilitate ``data structure-less'' programming, program
specification cannot rely on (existing) data structures and addressing
schemes, while still data has to be referenced. This is accomplished by
labeling each individual data item by a unique token. This approach is
similar to the approach taken in dataflow computing, as described
in~\cite{dennis-1974}, where tokens were used to match operands with each other
for each computational step.  Also, tokens can be seen in conventional
computing as the physical memory addresses of data stored in memory. In the
\emph{forelem} approach, these tokens do not necessarily imply physical
addresses or data items, etc., but are merely used to refer to elementary
data tuples which are being combined with other uniquely referenced data
fields. So, the basic ``data structure'' in \emph{forelem} consists of
tuples of the form:
\begin{equation*}
\langle token, data\rangle
\end{equation*}

\noindent
The whole approach now relies on the compilation framework to identify
different combinations of these elementary data tuples, thereby generating
more complex data structures. As such program specification is kept to a
minimum, only specifying which data fields are combined with
each other. In the \emph{forelem} framework, the tuple above is even further
stripped down to a normal form by specifying the above tuple as:
\begin{equation*}
\langle token\rangle _{A}
\end{equation*}

\noindent
indicating the data fields as $A(token)$ with $A$ an invertible function
(for example an affine function) that is referred to as an address function.
In most (all) cases the token field will consist of integers and the
function $A$ will be an affine mapping from $\mathbb{Z} \rightarrow
\mathbb{Z}$, and $A(token)$ is a logical memory address\footnote{For more
formal treatment of address functions, see the next sections.}. Also, the
token $i$ can be used for more than one address function. So, next to
$\langle i \rangle _{A}$, we can have additional tuples $\langle i \rangle
_{B}$, etc. Also, next to single token tuples, we can have multidimensional
token tuples, for instance $\langle i, j\rangle _{E}$. These
multidimensional tokens can be used to represent relationships between
token fields. In this case, the address function $E(i, j): \mathbb{Z}^2
\rightarrow \mathbb{Z}$ represents one value, see below for some examples.

\subsection{Example Program Specifications}
In this subsection, we give a number of examples of how traditional data
structures can be translated to elementary tuples.

\subsubsection{Graph data structures}
Graphs traditionally are represented by edges and vertices. Various data
structures have been devised for storing graphs, among which pointer-linked
approaches. In the pointer-linked approach, typically each vertex is stored
in memory as a record, or structure. Within each vertex a list is kept of
outgoing edges, which are stored as memory pointers to the structures
corresponding to the destination vertices. In our framework, the graph can
be represented as follows:

\begin{itemize}
\item Tuple $\langle i \rangle _{V}$ with $V$ the identity
function, where $V(i)$ represents the value of vertex $i$.
\item Tuple $\langle i, j\rangle _{E}$ with $E(i, j)$ is the weight of
edge $i \rightarrow j$.
\end{itemize}

\noindent
For a tuple $\langle i, j \rangle$ the following statement computes a new
weight for edge $i \rightarrow j$ such that the weight equals the absolute
value of the differences of the value of vertex $i$ versus the value of
vertex $j$:
\begin{equation*}
E(i, j) = abs(V(i) - V(j))
\end{equation*}

\subsubsection{Sparse}
Sparse matrices can be represented easily by the tuple $\langle i, j\rangle
_{A}$ where $A(i, j)$ represents the value of the matrix element $i, j$.

\subsubsection{Databases}
Relational databases rely on data tables stored as records, where related
fields are stored within the same record. These records could be seen as
tuples themselves: $\langle \mathrm{field_1, field_2, field_3}\rangle$.
However, when these records are being mapped onto our framework, the
\emph{forelem} tuples have the form: $\langle i\rangle _{A_1}, \langle
i\rangle _{A_2}, \langle i\rangle _{A_3}, \dots,\langle i \rangle _{A_n}$.
In this case, the relationship between the fields is represented by
the same token $i$ being used for each field. This specification reflects
the normal form of representing a database record in \emph{forelem},
although also other forms could be envisioned, including the one used in
relational databases.

\subsubsection{Linked lists}
Linked lists are normally represented as a pointer-linked data structure.
Each element of the list is placed as a structure in memory. Within each
element a ``link'' to the next (or previous) element is represented using a
pointer to that particular element. In \emph{forelem} this will look like
$\langle i, j\rangle _{V}$, where $V(i, j)$ represents the value of the
linked list chain element $\langle i, j \rangle$. The chains are linked to
one another by the property that for a tuple $\langle i_1, j_1 \rangle$
there exists exactly one tuple $\langle i_2, j_2 \rangle$ such that $j_1 =
i_2$.

Next to these elementary tuple specifications, the \emph{forelem} framework
allows the possible token values to be defined as a set. So, for instance
for sparse matrices only these tokens $\langle i, j \rangle$ can be
specified as belonging to a set $T$ which in fact represent non-zero
entries. Then, loop structures iterating over these token tuples can be
specified as iterating of all tuples $t \in T$~\footnote{For a detailed
treatment, see Section~\ref{sec:forelem-intro}}. Allowing these discrete specifications of
tuple reservoirs, most of the intricacies of conventional computing that
try to minimize storage by only storing non-zero entries, or in case of
graphs non-complete graphs, or in case of database systems where records are
stored as a universal relation, can be avoided resulting in very clean
specifications.

So, in case there is an iterative structure (for loop or while loop), each
operand to be used in an iteration can be referenced by a token out of a set
of tokens $T$. In fact, all of the tokens used as reference by a single
iteration are grouped together as a tuple. As an example, let's look at a graph
with edges $(i, j)$ and let's assume we want to perform the above assignment
over all edges. Then this can be expressed by the following loop
structure, which is a simple loop visiting all tuples in a tuple reservoir $T$:

\small
\begin{alltt}
\textbf{forelem} (t; t \(\in\) T)
\{
  E(t) = abs(V(t.i) - V(t.j));
\}
\end{alltt}
\normalsize

\noindent
Note that even in the case if there is no assignment to $E(i, j)$, we still
need the same tuple space $T$ as we are referencing both $V(t.i), V(t.j)$ in
the same iteration. So, the iterative construction can be considered as a
walk through the memory where the same functions are performed on each
collection of operands where the operands are being referenced or specified
by successive tuples in a tuple reservoir $T$.

Next to the fact that computations are specified on a single value, the
framework also assumes that in iterative constructs all the iterations can
in principle be executed independently from each other, thus in parallel.
This principle will ensure maximal optimization (repackaging) or the single
values into complex data structures. Next to the \emph{forelem} iterative
construct, the framework also contains a \emph{whilelem} construct. For the
latter construct, each iteration is also executed independently of other
iterations, but also are repeatedly executed depending on stopping
conditions.

\subsection{Examples Of Compiler-Generated Codes}
As an example, let us turn to our previous example of linked lists. Linked
lists are traditionally represented as a $\langle\mathrm{value},
\mathrm{next\_pointer}\rangle$ construct, but as has been described above
they might also be represented as tuples $\langle i,j \rangle _{V}$ with the
values stored as a separate array $\mathrm{V}[]$.  In the previous example,
we have seen that a (premature) choice of a linked list data structure leads
to dependencies in program execution which are non essential. If we now take
the example of a sorted list, so for all $\langle\mathtt{i, j}\rangle _{V}:
V(i) \leq V(j)$, the linked list representation would seem appropriate.
However, if inserts in this sorted list are represented in \emph{forelem}
representation as follows:

\small
\begin{alltt}
init()
\{
  T = \(\emptyset\);
  first_element = 0;
  V(first_element) = MAX;
\}

insert_value(new_value)
\{
  T = T \(\cup\) \(\langle\)first_element - 1, first_element\(\rangle\);
  V(first_element - 1) = new_value;

  \textbf{whilelem} (t; t \(\in\) T)
  \{
    \textbf{if} (V(t.i) > V(t.j))
      swap(V(t.i), V(t.j));
  \}

  first_element = first_element - 1;
\}
\end{alltt}
\normalsize

\noindent
Please note that the above code exemplifies both principles of the tUPL
framework. First of all, all references to the values are obtained by
$\langle t.i \rangle _{V}, \langle t.j \rangle _{V}$ as well as iterations
which can be executed independently from any other iterations. So, for
instance an execution order could be $\langle 1, 2 \rangle, \langle 3,
4\rangle , \langle 3, 2 \rangle , \langle 1, 4 \rangle, \dots$. This
potentially can result into an infinite loop, therefore the definition of
the \emph{whilelem} loop is such that all iterations can be executed
independently, assuming that just scheduling~\cite{lehmann-1981} is applied,
meaning that each of the $n$ iterations
will get a $100\% / n$ share of the total available CPU time. So in this
case, we have a iterative construct which addresses every single value
separately and independently allowing compiler techniques to be fully
exploited.

For instance, the compiler can affect the structure and organization of the
data to automatically generate data structures. Within the program
specifications no data structures are used and only the essential relations
between tokens and data items are specified. The compilation framework has
full freedom in generating a data structure that adheres to the specified
relations. To demonstrate the capabilities of the framework, we now show
several codes that can be automatically generated by the compiler from the
above specification of an insertion algorithm.

\subsubsection{Unordered Linked List Storage of Tuples}
The data storage of the example insertion algorithm is specified as being
tuples $\langle i, j \rangle _{V}$ with $V(i)$ representing the values. The
compiler can ``localize'' the tuples into tuples of the form
$\langle i, V(i), j \rangle$ and subsequently generate a data structure to
store these tuples. For instance, the tuples can be stored in a linked list
or array. For the former case, this results in the following code to be
generated:

\small
\begin{alltt}
changed = TRUE;
\textbf{while} (changed)
\{
  changed = FALSE;

  record = T.head;
  \textbf{while} (record != NULL)
  \{
    tmp = T.head;
    \textbf{while} (tmp != NULL)
    \{
      \textbf{if} (tmp->i == record->j)
        \textbf{break};
      tmp = tmp->j;
    \}
    assert(tmp != NULL);

    \textbf{if} (record->v > tmp->v)
    \{
      x = record->v;
      record->v = tmp->v;
      tmp->v = x;

      changed = TRUE;
    \}
    record = record->j;
  \}
\}
\end{alltt}
\normalsize

\noindent
Naturally, more sophisticated algorithms and/or indexing structures can be
used to obtain better performance, see also further on in this section.

\subsubsection{Array Ordered By Tuple Field Values}
The compiler can also choose to generate a data structure that uses an array
of structures, with the array elements stored in a particular order. With
$i$ being unique integer values, the elements can be stored according to
their $i$ value. As a consequence, the $i$ values can be used as subscript
into this array to access the corresponding tuple and because of this the
$i$ values do itself not have to be stored as part of the token in memory,
saving memory. The following code is generated for such a data structure:

\small
\begin{alltt}
changed = TRUE;
\textbf{while} (changed)
\{
  changed = FALSE;
  \textbf{for} (int i = 0; i < N; i++)
  \{
    \textbf{if} (T[i].V > T[T[i].J].V)
    \{
      x = T[i].V;
      T[i].V = T[T[i]J].V;
      T[T[i].J].V = x;

      changed = TRUE;
    \}
  \}
\}
\end{alltt}
\normalsize

\subsubsection{Linked List Ordered By Tuple Field Values}
Instead of an array subscript like above, $i$ can also represent a logical
memory address. In this case, a different code is generated:

\small
\begin{alltt}
changed = TRUE;
\textbf{while} (changed)
\{
  changed = FALSE;
  record = T.head;
  \textbf{while} (record != NULL)
  \{
    \textbf{if} (record->v > record->j->v)
    \{
      x = record->v;
      record->v = record->j->v;
      record->j->v = x;

      changed = TRUE;
    \}
     record = record->j;
  \}
\}
\end{alltt}
\normalsize

\subsubsection{Swapping Tuple Fields Instead of Data Fields}
In the above examples, the values $V(i)$ and $V(j)$ were swapped. Instead,
the compiler can also choose to modify the $i$ and $j$ tuple fields.
Because modifying tuple fields changes the semantics of the tuples stored in
the tuple reservoir, this is only allowed as a special operation that
substitutes values in all tuples contained in the tuple reservoir, referred
to as Global Substitution. So, in this case a swap of tuple fields is
carried out by by substituting $i, j$ for $j, i$ in all tuples.  Note that
this has the same effect at clearing the condition $V(t.i) > V(t.j)$ as
swapping $V(t.i)$ and $V(t.j)$. This leads the compiler to automatically
generate the following code, where the link chains $j$ are substituted (note
that because $i$ is not stored, it does not have to be substituted):

\small
\begin{alltt}
substitute(i, j)
\{
  record = start;
  \textbf{while} (record != NULL)
  \{
    tmp = record->j;
    if (record->j == i) record->j = j;
    else if (record->j == j) record->j = i;
    record = tmp;
  \}

  if (start == i) start = j;
  else if (start == j) start = i;
\}

changed = TRUE;
\textbf{while} (changed)
\{
  changed = FALSE;
  record = start;
  \textbf{while} (record != NULL)
  \{
    \textbf{if} (record->v > record->j->v)
    \{
      substitute(record, record->j);

      changed = TRUE;
    \}
    record = record->j;
  \}
\}
\end{alltt}
\normalsize

\subsubsection{Storage Ordered By Data Values}
One can also consider data to be ordered based on the value of the data
items $V(i)$, rather than the tuple field $i$. When new tuples and data are
inserted, the tuples and associated data values are automatically placed at
the correct position. The compiler accomplishes this by storing tokens and
data items in a data structure that is kept sorted and allows for quick
inserts, such as for instance sorted tree indexes used by database systems.
Due to the use of this data structure, the compiler is ensured that all
$V()$ values are stored in ascending order. Because of this, the compiler
will deduce that the if-condition will never be executed when $V(i) \leq
V(j)$ always holds. Clearly, this is the case when $i, j, ...$ are also
numbered in ascending order, so the compiler can ensure this property by
simply renumbering all $i$ and $j$ fields according to the order in which
$V(i)$ values are stored after each insert in the tuple
reservoir.

\subsubsection{Automatically Generating Multi-Dimensional Storage}
\begin{figure}
\begin{center}
\includegraphics[scale=0.42]{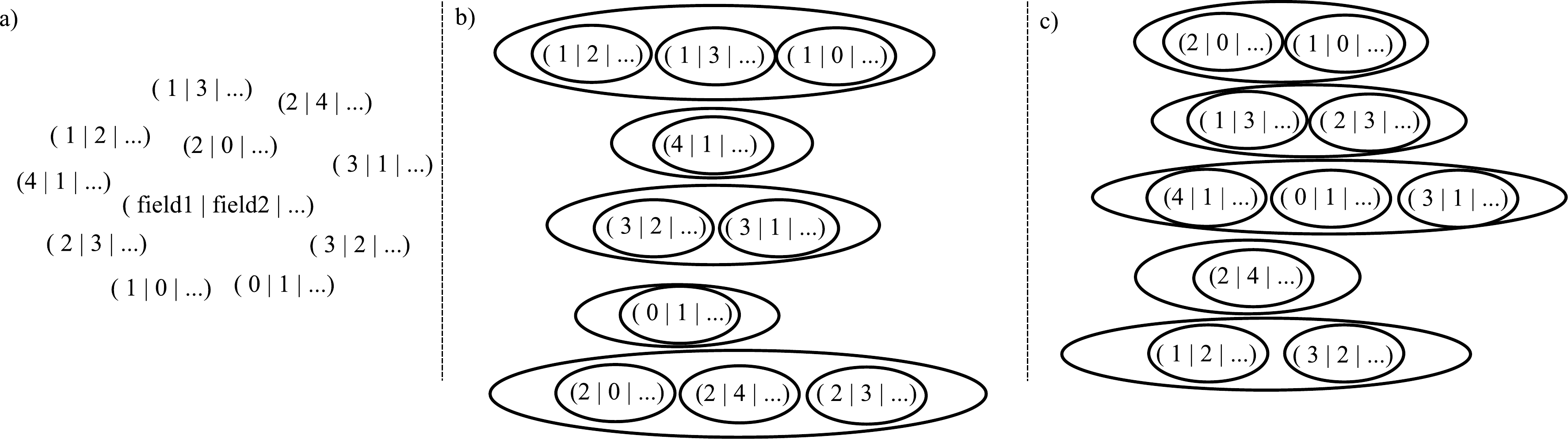}
\caption{a) Shows an initial, unordered, configuration of tuples. b) Tuples
are grouped based on \emph{i} (outer) and \emph{j} (inner). c)
Tuples are grouped based on \emph{i} (outer) and \emph{j}
(inner).}
\label{fig:groupings}
\end{center}
\end{figure}

Tuples of the form $\langle i, j \rangle$ could also be interpreted as
non-zero entries of a sparse matrix $\langle row, col \rangle$. In the case
of a linked list as is discussed here, this results in a sparse matrix with
1 non-zero entry per row. Within this two-dimensional structure, tuples can
be grouped in different ways, column or row-wise. Figure~\ref{fig:groupings}
illustrates how tuples are grouped according to different orders of tuple
fields.

From such a grouping of the tuples, different storage formats can be
automatically deduced. This includes established formats such as ITPACK and
JDS that were up till now only deduced by hand (see also
Section~\ref{sec:sparse-datagen}). In both cases matrix
elements (tuples) are grouped per row and matrix values and corresponding
column indices are stored separately. Through analysis the compiler will
establish that only a single entry per row is stored, causing a variant of
ITPACK to be generated that uses vector-storage instead of a two-dimensional
storage, resulting in the code:

\small
\begin{alltt}
changed = TRUE;
\textbf{while} (changed)
\{
  changed = FALSE;
  \textbf{for} (int i = 0; i < Nrows; i++)
  \{
    \textbf{if} (V[i] > V[J[i]])
    \{
      swap(V[i], V[J[i]]);

      changed = TRUE;
    \}
  \}
\}
\end{alltt}
\normalsize

\subsubsection{Automatic Generation of Sort Algorithms}
The function \texttt{insert\_value} can in fact be seen as a building block
of a sorting algorithm. The compiler can choose to delay the immediate
execution of the \emph{whilelem} loop and instead buffer insertions to
\verb!T! to be sorted all at once:

\small
\begin{alltt}
insert_value(new_value)
\{
  T = T \(\cup\) \(\langle\)first_element - 1, first_element\(\rangle\);
  V(first_element - 1) = new_value;
  first_element = first_element - 1;
\}

  /* multiple invocations of insert_value */
  insert_value(value1);
  insert_value(value2);
  ...
  insert_value(valuen);

  \textbf{whilelem} (t; t \(\in\) T)
  \{
    \textbf{if} (V(t.i) > V(t.j))
      swap(V(t.i), V(t.j));
  \}
\end{alltt}
\normalsize

\noindent
The compiler can apply multiple transformations to improve the performance
of the \emph{whilelem} loop. Because no iteration order is defined by the
loop structure, the compiler is free to select a suitable execution strategy
for the execution of this loop. One of these strategies is ``levelization'',
which organizes tuples into groups that have no dependencies between each
other, such that the groups can be processed in parallel. Tuples that
introduce a dependency between these groups can simply be excluded from
being processed, given that the tuples get a chance to the processed after
the parallel processing of the groups. Levelizations can be computed at
run-time (dynamic), but also static levelizations can be used.

An example levelization is one which increases the size of the group at
every iteration, leading to an execution strategy that is also used in the
Merge Sort algorithm\footnote{This strategy also bears similarity to the
``pointer jumping'' technique used by programmers to manually improve the
performance of parallel algorithms.}. When this levelization is applied,
this leads to the following code:

\small
\begin{alltt}
\textbf{for} (int i = 1; i < L; i++)
  \textbf{whilelem} (t; t \(\in\) T.i[not in \(n \times 2\sp{i}\)])
  \{
    \textbf{if} (V(t.i) > V(t.j))
      swap(V(t.i), V(t.j));
  \}
\end{alltt}
\normalsize

\noindent
It is obvious that the listed transformation areas unlock a myriad of
implementation choices.

\bigskip\noindent
As for linked lists the same techniques can be used for optimizing
applications in database systems, graph computations, etc., relying on a
very simple tuple specification of the computation and by employing a
compilation framework to generate the actual data structures, and by doing
so the actual choice of data structure is not a prerequisite for specifying
programs thereby alleviating the role of the programmer. In fact, we argue
that a choice of data structures in conventional programming is nothing more
than an initial way of representing computation and as such are not
essential for generating an executable program. Whilst there have been many
approaches in the past which had the same goal, most of
these approaches failed because of existing compiler techniques not being
able to support these approaches in an adequate fashion so that efficient,
high-performing codes could be generated. In our case, the goal is not to
just go for a higher level of abstraction, but to use the tuple
specification to directly enable compiler technology to generate efficient
codes. So, whereas there might be similarity to an approach like for
instance Linda, the underlying principle is very different and we would not
advocate our approach as an alternative to conventional computing, (but
merely as a tool to be used to obtain high performing
codes\footnote{Although approaches with functional programming and Linda
addressed the fact that the parallelization of programs is more natural and
can be effectively exploited, backend compilation / code generation was
never effectively exploited.}).

As can be seen from this example, the use of this different means of program
specification puts the compiler back into the game and enables the compiler
to perform a very large number of program transformations and optimizations
that reach significantly beyond transformations applied by contemporary
compilers. In this paper, as a first step we will concentrate on sparse
matrix computations and the compiler-based generation of data structures. We
show that the compiler is capable of generating various data structures that
differ in data grouping and ordering, and that this way the compiler is able
to generate high performance codes.

\section{The Forelem framework}
\label{sec:forelem-intro}
Our approach for the automatic generation of data structures is based on the
\emph{forelem} framework~\cite{rietveld-2013-cpc}. Therefore, in this
section the basics of the \emph{forelem} framework are briefly introduced.
The \emph{forelem} framework provides a different scheme to specify programs
in which the specification of fixed iteration order and fixed specification
of data structures is explicitly avoided. In this framework, tuples are used
as elementary data structure and traditional data structures need to be
`disassembled' into tuples. Programs are composed by writing manipulations
of sets of tuples. In fact, central to this scheme is the
\emph{forelem} loop construct. Each \emph{forelem} loop iterates over a
specific (subset of a) tuple reservoir and each selected tuple will be
visited exactly once. Note that it is not determined how the tuples are
stored, as actual data structures will be generated during the optimization
process, see the next Section.

The tuples are used as tokens to retrieve additional data fields belonging
to this tuple from memory. As an example, consider tuples with fields
\emph{field1} and \emph{field2} collected in a tuple reservoir \verb`T`.
The following \emph{forelem} loop iterates all tuples of \verb`T` and
accesses the corresponding values using the address function \verb`A`:

\small
\begin{alltt}
\textbf{forelem} (t; t \(\in\) T)
  ... A(t.field1) ...
\end{alltt}
\normalsize

\noindent
Although the \emph{forelem} loop appears to be very similar to a
\emph{foreach} loop that exists in many common programming languages,
\emph{forelem} loops distinguish themselves with the use of tuple reservoirs
to avoid fixed specification of data structures and the explicitly undefined
iteration order of the tuples. Additionally, a \emph{whilelem} loop is
defined which is also unordered and may repeatedly visit particular tuples
depending on stopping conditions. For \emph{forelem} loops, the exact
semantics of the iteration of the tuple reservoirs is to be determined
in the course of the optimization process. By specifying programs in this
form, the compiler is no longer obstructed by fixed specifications of codes
and data structures that it cannot violate. As a result, this allows the
compiler to go beyond the optimization of solely control flow and will also
target the way data is organized and accessed.

Using a special syntax, it is possible to narrow down the tuples of a tuple
reservoir that are iterated. In fact, subsets of the tuple reservoir can be
selected based on particular values or value ranges of the named fields.
For example, \verb!T.field2[k]! will select only those tuples in \verb`T`
for which \emph{field2} has value \verb`k`.
This is expressed mathematically as follows:
\begin{equation*}
\mathtt{T.field2[k]} \equiv \{ \mathtt{t} \mid \mathtt{t} \in \mathtt{T}
\wedge \mathtt{t.field2} = \mathtt{k}\}
\end{equation*}
\noindent
So, to only iterate tuples of \verb!T! in which the value of \emph{field2} is
$10$, the following \emph{forelem} loop is used:

\small
\begin{alltt}
\textbf{forelem} (t; t \(\in\) T.field2[10])
  ... A(t.field1) ...
\end{alltt}
\normalsize

\noindent
Note that \verb!T.field2[10]! is not expressed more explicitly as the
exact execution of the loop will be determined during the optimization process.
This particular subset of the tuple reservoir might be explicitly generated
at either compile- or run-time, it might be contained with other
\emph{forelem} loops that iterate the same tuple reservoir, or might be
eliminated.
Alternatively, during the optimization process it may be decided to create a
variant of \verb!T! only containing the tuples to be iterated.

More sophisticated selections are possible, such as having conditions on multiple
fields, in this case on \emph{field1} and \emph{field2}:
\begin{equation*}
\mathtt{T.}(\mathtt{field1},\mathtt{field2})[(\mathtt{k}_1,\mathtt{k}_2)]
\equiv \\
\{\mathtt{t} \mid \mathtt{t} \in \mathtt{T} \wedge \mathtt{t.field1} =
\mathtt{k}_1 \wedge \mathtt{t.field2} = \mathtt{k}_2\}
\end{equation*}
\noindent
Instead of a constant value, the values $\mathtt{k}_{n}$ can also be
references to values from a tuple from another reservoir (e.g., in nested
loops).
To select values $\mathtt{field1} > 10$ an interval is
used: $(10, \infty)$.

\section{Transformations For Automatic Data Structure Generation}
\label{sec:transformations}
The automatic generation of data structures is directed by the application
of different code transformations. In this section we describe several
transformations defined in the \emph{forelem} framework, that
are used to automatically derive different implementations of a loop
structure and accompanying data structure. The subsequent section,
Section~\ref{sec:ds-gen}, discusses how different compositions of these
transformations lead to different data structures.

\subsection{Orthogonalization}
\label{sec:orthogonalization}
In \emph{forelem} loops, the iteration order of the tuples in a tuple
reservoir is explicitly undefined.
In this section, the
\emph{orthogonalization} transformation is introduced, which makes it
possible to impose a certain order in which the tuple reservoir is iterated. This is
achieved by partitioning the accesses to the reservoir based on the values of
one or more tuple fields. The orthogonalization transformation is used to
control the order in which data is accessed as a preparatory step to
Materialization, which is discussed in the next section.

Let \verb!T! be a tuple reservoir with tuples with fields $\mathtt{field1},
\mathtt{field2}, \dots, \mathtt{fieldn}$ collected in a tuple reservoir
\verb!T!, and the loop:

\small
\begin{alltt}
\textbf{forelem} (t; t \(\in\) T)
  ... A(t) ...
\end{alltt}
\normalsize

\noindent
In this loop, the tuples of \verb!T! can be iterated in any order. As an
example, assume an iteration order is to be imposed on \verb!T! such that
tuples \verb!T! are accessed in blocks with equal values for \verb!field1!.
The orthogonalization transformation is carried out to achieve this,
resulting in the following loop nest:

\small
\begin{alltt}
\textbf{forelem} (i; i \(\in\) T.field1)
  \textbf{forelem} (t; t \(\in\) T.field1[i])
    ... A(t) ...
\end{alltt}
\normalsize

\noindent
\verb!T.field1! in the other loop denotes a reservoir of singleton tuples
representing all values for field \verb!field1! in the reservoir \verb!T!.
So, the iteration space of the outer
loop consists out of every value of \verb!field1! in \verb!T!.
The original loop iterates all tuples of \verb!T!. The transformed loop nest
will, for every value of \verb!field1!, iterate all tuples of \verb!T! for
which \verb!field1! equals this value. As a result, the transformed loop also
iterates all tuples of \verb!T!.  Application of the orthogonalization
transformation is not limited to a single field. An example of
orthogonalization on two fields is:

\small
\begin{alltt}
\textbf{forelem} (i; i \(\in\) T.field1)
  \textbf{forelem} (j; j \(\in\) T.field2)
    \textbf{forelem} (t; t \(\in\) T.(field1,field2)[(i,j)])
      ... A(t) ...
\end{alltt}
\normalsize

The outer loops that are introduced by the orthogonalization transformation
iterate all values of a given tuple field. If it is possible to express this
range of values as a subset of the natural numbers, i.e. $\mathtt{T.field1}
\subseteq \mathbb{N}$, the \emph{encapsulation} transformation can be applied,
which replaces the loop over all tuple field values with a loop over a
subset of the natural numbers.  With the encapsulation transformation, a
loop

\small
\begin{alltt}
\textbf{forelem} (i; i \(\in\) T.field1)
\end{alltt}
\normalsize

\noindent
where $\mathtt{T.field1} = \{1, 2, 6, 7, 8, 10\}$, is replaced with:

\small
\begin{alltt}
\textbf{forelem} (i; i \(\in \mathbb{N}\sb{10}\))
\end{alltt}
\normalsize

\noindent
with $\mathbb{N}_{10} = \{1,  ..., 10\}$. In the encapsulated loop, the
values $3, 4, 5, 9$ will be iterated, but note that no tuple will exist
where \verb!field1! equals any of these values. As a result, the inner loop
is not executed for these values, maintaining the iteration space of the
original loop.

\subsection{Materialization}
\label{sec:materialization}
The materialization transformation
materializes the tuples iterated by a \emph{forelem} loop from a
particular (subset of a) tuple reservoir
to a sequence in which the data is represented in
consecutive order and is accessed with integer subscripts. Although this can
be seen as a simple normalization operation, it is an important enabling
step that allows the compiler to address and modify the order of data access
to this sequence. In fact, by materialization an execution order, or a
sequence, is determined for elements iterated by an inner loop.  (In the
case of nested loops, orthogonalization determines the order of the
outermost loop).  After two forms of materialization have been introduced, a
number of transformations targeting the order in which data access takes
place will be described.

A distinction is made between loop-independent and loop-dependent
materialization. In loop-independent materialization, conditions on the
tuple reservoir of the loop to be materialized are not dependent on one of the
outer loops. Materialization will result in a single sequence. In
loop-dependent materialization, sequences are nested and an additional
nesting level is added for each dependent loop. Both cases of
materialization will now be discussed in turn. Throughout the discussion we
will use the C one-dimensional array notation to denote sequences, the
two-dimensional array notation to denote sequences of sequences
and so on. This array notation is purely symbolic and does not imply that
the data is actually stored as an array in memory.

\subsubsection{Loop Independent Materialization}
We first consider loop-independent materialization. The following loop
iterates all tuples of \verb!T! whose field equals a value \verb!X!:

\small
\begin{alltt}
\textbf{forelem} (t; t \(\in\) T.field[X])
  ... A(t) ...
\end{alltt}
\normalsize

\noindent
To be able to determine which tuples of \verb!T!, and which values in
\verb!A!, to access, a condition on
the tuple reservoir is used.
In fact, this is an indirection level. This indirection can be removed
by materializing the index into the tuple space as a sequence \verb!PA! which
only contains the entries of \verb!A! that should be visited by this loop.
This results in:

\small
\begin{alltt}
\textbf{forelem} (i; i \(\in \mathbb{N}*\))
  ...  PA[i]  ...
\end{alltt}
\normalsize

\noindent
with $\mathbb{N}* = [0, \mathtt{|PA|-1}]$. The sequence \verb!PA! only contains
elements from \verb!A! for which the condition \verb!t.field == X! holds.
The compiler is now enabled to address the order in which the data in
\verb!PA! is accessed, while the execution order of the loop is not
specified.  For example, using the transformations that can be applied on
the materialization form the compiler can determine to put entries in
\verb!PA! in a specific order. The loop control is selected at the
concretization stage, where the compiler can ensure the loop control for the
loop will iterate the items of \verb!PA! consecutively.  For the general
definition of loop-independent materialization, consider a loop iterating a
sparse structure \verb!A! with a reservoir \verb!T!:

\small
\begin{alltt}
\textbf{forelem} (t; t \(\in\) T)
  ...  A(t)  ...
\end{alltt}
\normalsize

\noindent
which is transformed to:

\small
\begin{alltt}
\textbf{forelem} (i; i \(\in \mathbb{N}*\))
  ...  PA[i]  ...
\end{alltt}
\normalsize

\noindent
with $\mathbb{N}* = [0, \mathtt{|PA|-1}]$. This transformation materializes
the sparse structure to a single sequence \verb!PA!.

\subsubsection{Loop Dependent Materialization}
If a loop to be materialized is contained in a loop nest and the conditions
on its tuple reservoir have a dependency on another loop, then the above described
loop-independent materialization cannot be applied. Instead, loop-dependent
materialization must be used, which is described in this section. Because
loop-dependent materialization will result in nested sequences,
this results in more opportunities for the compiler to address and modify
the order of data access to these sequences.  In general, a loop-dependent
materialization has the form:

\small
\begin{alltt}
\textbf{forelem} (i; i \(\in \mathbb{N}\sb{o}\))
  ...
    \textbf{forelem} (n; n \(\in \mathbb{N}\sb{t}\))
      \textbf{forelem} (t; t \(\in\) T.(field\(\sb{i}\), ...,field\(\sb{n}\))[(i,...,n)])
         ... A(t) ...
\end{alltt}
\normalsize

\noindent
where $\mathbb{N}_{o} = \{1, 2,  ..., o\}$. In the encapsulated loop, the
tuple reservoir iterated in the inner loop has a dependency on one or more of
the outer loops. The elements that are visited in \verb!A! are materialized to an iteration
of nested sequences \verb!PA!, in which each loop-dependent condition is
represented as an additional nesting level in \verb!PA!. The sequence
\verb!PA! only contains these items that were iterated using the
tuples from the original tuple reservoir \verb!T!:

\small
\begin{alltt}
\textbf{forelem} (i; i \(\in \mathbb{N}\sb{o}\))
  ...
    \textbf{forelem} (n; n \(\in \mathbb{N}\sb{t}\))
      \textbf{forelem} (p; p \(\in\mathbb{N}*\))
        ... PA[i]...[n][p] ...
\end{alltt}
\normalsize

\noindent
with $\mathbb{N}* = [0, \mathtt{|PA[i]...[n]|-1}]$. After this transformation,
\verb!PA! only contains entries that satisfy the conditions of the original
conditions on the tuple reservoir.  The dimensions of the materialized
sequence correspond to the original conditions and thus to the loops on
which the condition depended.

\subsection{Transformations on the Materialized Form}
After a \emph{forelem} loop has been put in a materialized form, the data to
be processed has been put in a sequence in consecutive order and is accessed
with integer subscripts. At this stage, the compiler can modify the exact order
of data access to these sequences and how this data is stored. In this section
a number of transformations are described that affect the storage of the
data processed by a loop nest.

\subsubsection{Horizontal Iteration Space Reduction}
The aim of Horizontal Iteration Space Reduction is to reduce unused fields
from the tuples.  In fact, it is possible to perform this transformation
before the materialization stage.
Formally, the transformation is defined as follows.
Let \verb!T! be a tuple reservoir containing tuples with fields \verb!field1!, \verb!field2!,
\verb!field3!, \verb!field4!,
and \verb!C! a list of condition fields
$\mathtt{C} \subset (\mathtt{field1}{\;}\mathtt{field2})$ and \verb!V! a list
of values. Consider the loop nest:

\small
\begin{alltt}
\textbf{forelem} (t; t \(\in\) T.C[V])
  ... t.field1  ...  t.field2  ...
\end{alltt}
\normalsize

\noindent
We define a new reservoir $\mathtt{T'} \subseteq \mathtt{T}$ with fields
\verb!field1!, \verb!field2!  and replace the use of $\mathtt{T}$ with
$\mathtt{T'}$ in the loop.

\subsubsection{Tuple splitting}
Before materialization, token tuples are used to refer to elementary data
tuples.  By default, the sequence that is the result of the materialization
operation is a sequence of data tuples or structures, accessible with
integer subscripts. In some cases, it is more efficient to use a structure
of sequences, i.e., the structures are split~\cite{spek-2008,curial-2008}.
Within the \emph{forelem} framework this is defined as the \emph{structure
splitting} transformation. Consider the materialized loop nest:

\small
\begin{alltt}
\textbf{forelem} (i; i \(\in \mathbb{N}\sb{m}\))
  \textbf{forelem} (k; k \(\in \mathbb{N}*\))
    ... PA[i][k].value ...
\end{alltt}
\normalsize

\noindent
Structure splitting will modify the data storage of the sequence and convert
the data accesses in the loop to:

\small
\begin{alltt}
\textbf{forelem} (i; i \(\in \mathbb{N}\sb{m}\))
  \textbf{forelem} (k; k \(\in \mathbb{N}*\))
    ... PA.value[i][k] ...
\end{alltt}
\normalsize

\subsubsection{$\mathbb{N}*$ materialization}
Materialized loops use the $\mathbb{N}*$ set as the set of integer
subscripts to access the materialized sequence. How exactly these integer
subscripts are stored is initially encapsulated within $\mathbb{N}*$ and can
be made explicit using $\mathbb{N}*$ materialization.
Consider the following loop, the result of a materialization to \verb!PA!:

\small
\begin{alltt}
\textbf{forelem} (i; i \(\in \mathbb{N}\sb{m}\))
  \textbf{forelem} (k; k \(\in \mathbb{N}*\))
    ... PA[i][k] ...
\end{alltt}
\normalsize

\noindent
As a prerequisite for the final code generation stage, $\mathbb{N}*$ must be
made explicit. This can be achieved by converting $\mathbb{N}*$ to a set
\verb!PA_len!. There are different means in which this set can be defined.
The first is to define the set as follows:

\small
\begin{alltt}
  PA\_len[q] = max(len(PA[q]))
\end{alltt}
\normalsize

\noindent
in which case all \verb!PA_len[q]! values are the same and a single set
containing integers up to the
maximal value can be stored for this loop nest. Padding is inserted
in the sequence \verb!PA! for the values \verb!PA[i][k]! with
\verb!k >= PA_len[i]!.  The second way to create this sequence is to avoid inserting
padding in \verb!PA!. In this case \verb!PA_len[q] = len(PA[i])!.
Regardless of which implementation is chosen, the resulting loop after
$\mathbb{N}*$ materialization is:

\small
\begin{alltt}
\textbf{forelem} (i; i \(\in \mathbb{N}\sb{m}\))
  \textbf{forelem} (k; k \(\in\) PA\_len[i])
    ... PA[i][k] ...
\end{alltt}
\normalsize

\noindent
Note that in this loop the iteration order is still undefined. Only
$\mathbb{N}* = [0, \mathbb{N*}-1]$ has been replaced with
$\mathtt{PA\_len[i]} = [0, \mathtt{PA\_len[i]}-1]$.
In a subsequent concretization step (see Section~\ref{sec:concretization} the iteration order will be
determined. For example, the loop:

\small
\begin{alltt}
  \textbf{forelem} (k; k \(\in\) PA\_len[i])
\end{alltt}
\normalsize

\noindent
is concretized to:

\small
\begin{alltt}
   \textbf{for} (k = 0; k < PA\_len[i]; k++)
\end{alltt}
\normalsize

\subsubsection{$\mathbb{N}*$ sorting}
In case of loop-dependent materialization, $\mathbb{N}*$ encapsulates the
sets of integer subscripts used for iteration of the inner loop.
These sets are ordered irrespective of their cardinality. If
the loop is to be parallelized, it is beneficial if the work is divided
into blocks with evenly sized values for \verb!PA_len! (after $\mathbb{N}*$
materialization). One way to achieve this
is by imposing an order on the iteration of $\mathbb{N}*$.

The aim of $\mathbb{N}*$ sorting is to find an order of the iterator values
\verb!i! such that the value of $\mathbb{N}*$ decreases with subsequent
iterations of the outer loop on \verb!i!

\small
\begin{alltt}
\textbf{forelem} (i; i \(\in \mathbb{N}\sb{m}\))
  \textbf{forelem} (k; k \(\in \mathbb{N}*\))
    ... PA[i][k] ...
\end{alltt}
\normalsize

\noindent
Consider that $\mathbb{N}* = [0, \mathtt{len(PA[i])}-1]$. The goal is to iterate
through $\mathbb{N}_{m}$, such that \verb!len(PA[i])! decreases. Let
$\mathtt{perm}(\mathbb{N}_m)$ store the permutation of $\mathbb{N}_m$ for
which this holds. Then, the loop is transformed to:

\small
\begin{alltt}
\textbf{forelem} (i; i \(\in\) perm(\(\mathbb{N}\sb{m}\)))
  \textbf{forelem} (k; k \(\in \mathbb{N}*\))
    ... PA[i][k] ...
\end{alltt}
\normalsize

\noindent
Note that this will affect the order of the data \verb!PA!, which will be
put in the corresponding sorted order at the concretization stage.

\subsubsection{Dimensionality Reduction}
Loop-dependent materialization results in a nested sequence by
default. If this sequence is concretized as a multi-dimensional array, padding
may have to be inserted for the uneven lengths of the rows. It is possible
to avoid the introduction of this padding by storing the sequences back to
back.  This reduces the nesting depth of the materialized sequence and the
dimensionality of the multi-dimensional array that could be the result of
concretization.  Consider
the loop nest:

\small
\begin{alltt}
\textbf{forelem} (i; i \(\in \mathbb{N}\sb{m}\))
  \textbf{for} (k = 0; k < PA\_len[i]; k++)
     ... PA[i][k] ...
\end{alltt}
\normalsize

\noindent
to reduce the dimensionality of the materialized sequence \verb!PA! by one,
this is transformed into:

\small
\begin{alltt}
\textbf{forelem} (i; i \(\in \mathbb{N}\sb{m}\))
  \textbf{for} (k = PA\_ptr[i]; k < PA\_ptr[i+1]; k++)
     ... PA[k] ...
\end{alltt}
\normalsize

\noindent
Based on the \verb!PA_len! array, a new \verb!PA_ptr! array is introduced,
which keeps track of the start and end of each row in \verb!PA!.  Note that
the order of the iteration domain $[\mathtt{PA\_ptr[i]},
\mathtt{PA\_ptr[i+1]}-1]$ does not have to be defined and could be in any order.

%
%

\section{Automatic Data Structure Generation}
\label{sec:ds-gen}
We now demonstrate that when materialization is integrated with other
transformations that are defined within the \emph{forelem} framework, many
different data storage formats can be generated for a particular problem.

\subsection{Consolidating Data Using Loop Collapse}
With Loop Collapse, two loops accessing different tuple reservoirs are
collapsed to a single loop that accesses a single, combined, tuple
reservoir. Given the following starting point:

\small
\begin{alltt}
\textbf{forelem} (t; t \(\in\) T)
  \textbf{forelem} (r; r \(\in\) R.b_field[t.a_field])
    ... A(t) ... B(r) ...
\end{alltt}
\normalsize

\noindent
which is transformed to:

\small
\begin{alltt}
\textbf{forelem} (t; t \(\in\) TxR.b_field[a_field])
  ... A(t) ... B(r) ...
\end{alltt}
\normalsize

\noindent
The special notation used in the tuple reservoir condition guarantees that only these tuples
\verb!t!
are iterated that satisfy the condition
\verb!t.b_field == t.a_field!. So, the cross product that is
iterated by the loop body of the loop nest is exactly the same before and
after the Loop Collapse transformation. If subsequently the
(loop-independent) materialization transformation is applied, then a
sequence \verb!PAxB! is created.  This sequence contains data that
was originally stored in the separate \verb!A! and \verb!B! structures. These
original structures have thus been disassembled and reassembled into a single
data structure \verb!PAxB!. The resulting loop structure is:

\small
\begin{alltt}
\textbf{forelem} (i; i \(\in \mathbb{N}\sb{\mathtt{AxB}}\))
  ... PAxB[i] ...
\end{alltt}
\normalsize

\subsection{Generating Different Data Groupings With Loop Interchange}
As the next example we will discuss the Loop Interchange
transformation~\cite{allen-1984}. Two different ways of applying the Loop
Interchange transformation are described, which when combined with
materialization give rise to different data structures. Let \verb!T! be a
tuple reservoir containing tuples with fields \verb!field1! and \verb!field2!.
Using these tuples, data tuples in \verb!A! are accessed.
Consider the following starting point:

\small
\begin{alltt}
\textbf{forelem} (t; t \(\in\) T)
  ... A(t) ...
\end{alltt}
\normalsize

\noindent
The tuples of \verb!T! are iterated in a fully undefined order. This can
seen in Figure~\ref{fig:groupings}a). Some order can be created using the
Orthogonalization transformation, to result in for instance:

\small
\begin{alltt}
\textbf{forelem} (i; i \(\in \mathbb{N}\sb{m}\))
  \textbf{forelem} (j; j \(\in \mathbb{N}\sb{n}\))
    \textbf{forelem} (t; t \(\in\) T.(field1,field2)[(i, j)])
      ... A(t) ...
\end{alltt}
\normalsize

\noindent
In this loop nest, the access pattern of the tuples in \verb!T! is
orthogonalized on two fields of \verb!T!. The outer loop visits groups of
tuples with equal values for \verb!field1!. The middle loop visits groups of
tuples with equal values for \verb!field2!, within a group of tuples of
equal values of \verb!field1!. The inner loop thus visits tuples with equal
values for both \verb!field1! and \verb!field2!. For the sake of simplicity,
we assume that at most one such tuple exists. So, all data in \verb!T!
(and in turn in \verb!A!) is
processed in groups of tuples with equal values for \verb!field1!. This is
depicted in Figure~\ref{fig:groupings}b). In this figure, each circle denotes a
group. The order in which these groups are visited and the order in which
the tuples within each group are visited is not defined.  When the two outer
loops are interchanged, this grouping order is changed:

\small
\begin{alltt}
\textbf{forelem} (j; j \(\in \mathbb{N}\sb{n}\))
  \textbf{forelem} (i; i \(\in \mathbb{N}\sb{m}\))
    \textbf{forelem} (t; t \(\in\) T.(field1,field2)[(i, j)])
      ... A(t) ...
\end{alltt}
\normalsize

\noindent
As a result, the outer loop now visits groups of tuples that have equal
values for \verb!field2! instead of \verb!field1!. This changes the way the
tuples of \verb!T! are grouped, as can be seen in
Figure~\ref{fig:groupings}c). So, Loop Interchange changes the way
tuples are grouped.

Now, consider the loop nest before Loop Interchange as described above. In
this loop nest, the orthogonalization on \verb!field2! can be undone,
resulting in:

\small
\begin{alltt}
\textbf{forelem} (i; i \(\in \mathbb{N}\sb{m}\))
  \textbf{forelem} (t; t \(\in\) T.field1[i])
    ... A(t) ...
\end{alltt}
\normalsize

\noindent
Subsequently, loop-dependent materialization of the inner loop leads to:

\small
\begin{alltt}
\textbf{forelem} (i; i \(\in \mathbb{N}\sb{m}\))
  \textbf{forelem} (k; k \(\in \mathbb{N}*\))
    ... PA[i][k] ...
\end{alltt}
\normalsize

\noindent
with $\mathbb{N}* = [0, \mathtt{|PA[i]|-1}]$.  The iteration of \verb!T! has
been materialized to an iteration of sequences \verb!PA[i]!, in
which each loop-dependent condition is represented as a separate
sequence in \verb!PA!. The resulting structure \verb!PA! only
contains these items that are iterated by the original tuple reservoir.
What was previously a group with undefined order, has now become a
sequence of items. So, for each outer iteration \verb!i!, tuples
\verb!PA[i][k]! with \verb!k = 1, 2, ...! are accessed. See
Figure~\ref{fig:matr}a). If Loop Interchange
is now applied then \verb!k!  becomes the outer loop, leading to:

\begin{figure}
\begin{center}
\includegraphics[scale=0.42]{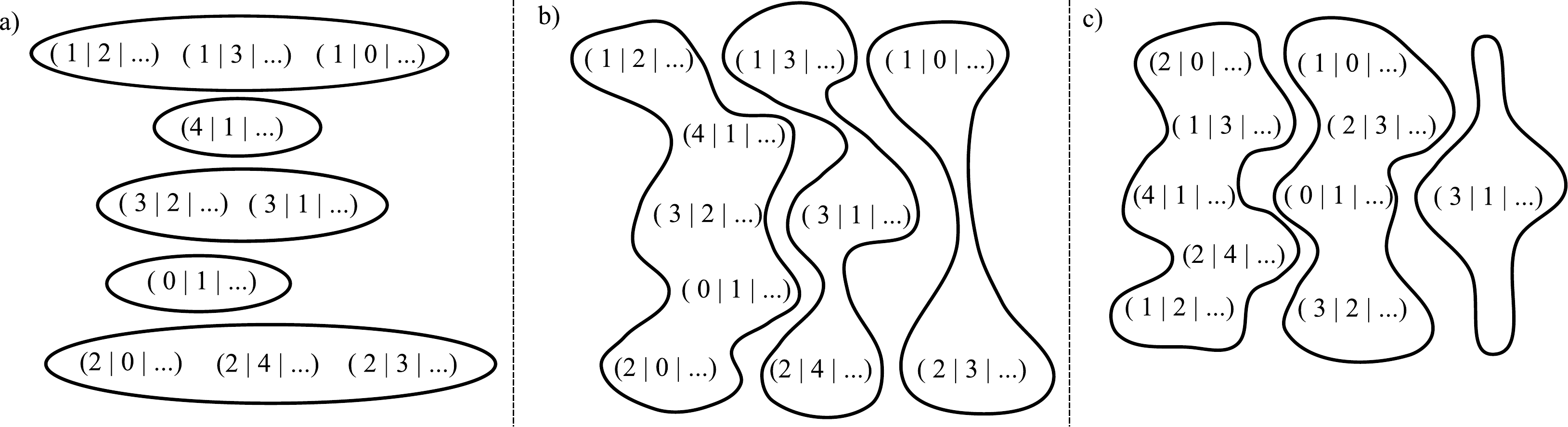}
\caption{a) Configuration of the tuples after loop-dependent materialization
of a loop with a condition on \emph{field1}. b) Grouping of the tuples after
Loop Interchange has been applied on the loop nest iteration the structure
a). c) The same as b), but using a loop nest with a condition on
\emph{field2} as starting point.}
\label{fig:matr}
\end{center}
\end{figure}

\small
\begin{alltt}
\textbf{forelem} (k; k \(\in \mathbb{N}*\))
  \textbf{forelem} (i; i \(\in \mathbb{N}\sb{m}\))
    ... PA[i][k] ...
\end{alltt}
\normalsize

\noindent
This results in an execution order where for each tuple position $k$ within
a sequence, all sequences \verb!PA[i]! are accessed. So, first all tuples at
the first positions in the sequences are accessed, then all second tuples,
and so on. See Figure~\ref{fig:matr}b). Note that this has led to a
grouping of the tuples that is different from the tuple organizations seen
in Figure~\ref{fig:groupings}. Figure~\ref{fig:matr}c) shows the result of
undoing the orthogonalization on \verb!field1!, followed by materialization
and loop interchange. Also in this case, compare with
Figure~\ref{fig:groupings}
and note that a different organization has been generated.

\subsection{Creating Data Partitions Using Loop Blocking}
To conclude this section, we describe how the Loop Blocking transformation
affects the automatic generation of data structures. Consider the following
simple orthogonalized loop nest:

\small
\begin{alltt}
\textbf{forelem} (i; i \(\in \mathbb{N}\sb{m}\))
  \textbf{forelem} (t; t \(\in\) T.field[i])
    ... A(t) ...
\end{alltt}
\normalsize

\noindent
To process the data in blocks of \emph{T.field} values the iteration space
$\mathbb{N}\sb{m}$ is partitioned into $x$ blocks:

\begin{equation*}
\mathbb{N}_m = \mathbb{N}_{[0, x)} \cup \mathbb{N}_{[x, 2x)} \cup ... \cup \mathbb{N}_{[(\mathbb{N}_m/x - 1)x, \mathbb{N}_m)}
\end{equation*}

\noindent
The loop blocking transformation adds an additional loop to iterate through
these partitions of $\mathbb{N}\sb{m}$. This results in the following loop
nest:

\small
\begin{alltt}
\textbf{forelem} (ii; ii \(\in \mathbb{N}\sb{m/x}\))
  \textbf{forelem} (i; i \(\in \mathbb{N}\sb{[ii{\cdot}x, (ii+1){\cdot}x)}\))
    \textbf{forelem} (t; t \(\in\) T.field[i])
      ... A(t) ...
\end{alltt}
\normalsize

\noindent
Note that the inner loop has a dependency on the middle loop,
which subsequently has a dependency on the outer loop. As a consequence,
when the inner loop is materialized, both the middle and inner loops will be
transformed.
Materialization will result in:

\small
\begin{alltt}
\textbf{forelem} (ii; ii \(\in \mathbb{N}\sb{m/x}\))
  \textbf{forelem} (i; i \(\in \mathbb{N}\sb{x}\))
    \textbf{forelem} (j; j \(\in \mathbb{N}*\))
      ... PA[ii][i][j] ...
\end{alltt}
\normalsize

\noindent
\verb!PA! is a symbolic data structure. As has been described, this
this symbolic array is later mapped to a physically allocated data
structure. At this point, it has not been decided that \verb!PA! will be
stored as a three-dimensional array. Rather, different sub-data structures
might be generated for \verb!PA[0][][]!, \verb!PA[1][][]!, and so on.

\begin{figure}
\begin{center}
\includegraphics[scale=0.75]{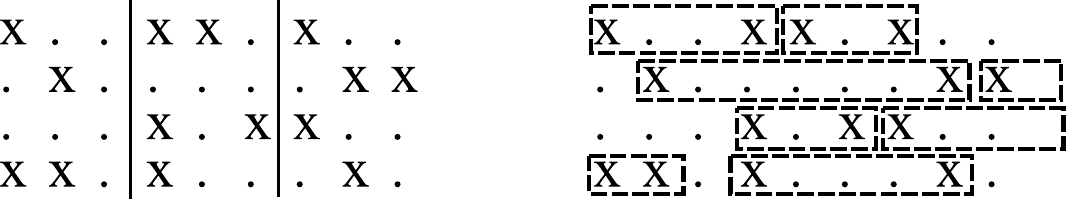}
\caption{Left: effect of blocking before Materialization. Right: effect of
blocking after Materialization. An $x$ denotes a nonzero element.}
\label{fig:matr-blocking}
\end{center}
\end{figure}

The partitions that are created by blocking on a dimension $\mathbb{N}_m$,
as we have just seen in the example, are shown in
Figure~\ref{fig:matr-blocking} on the left. In this case, loop blocking is
performed before Materialization and because of this partitioning is done
regardless of the tuples that exist within \verb!A!.  For example, it may be
the case that for a certain partition $\mathbb{N}_i \subset \mathbb{N}_m$,
no tuple exists in \verb!A! such that $\mathtt{field} \in \mathbb{N}_i$.
This may lead to blocks with highly irregular amounts of tuples. Consider
now that Materialization is performed first, followed by loop blocking. In
this case, only the tuples that are actually present in \verb!A! are
considered. Blocking is performed on the dimensions of the matrix containing
only the nonzeros. As a consequence, blocks of data that are to be processed
will be much more evenly filled, as is shown in
Figure~\ref{fig:matr-blocking} on the right.

\section{Case Study: Automatic Generation of Sparse Matrix Data Structures}
\label{sec:case-study}
In this section, we present a case study in which we apply the framework to automatically
generated data structures described in paper to sparse matrix routines.
We first describe how sparse computations can be expressed as operating on
tuples. These tuples are a natural result from disassembling the original
sparse matrix structures. The nonzero entries of a sparse matrix are
isolated from the other entries and treated as separate tuples.
Subsequently, we describe how from sparse computations expressed in this
manner many different routines and data structures can be instantiated
automatically using the code transformations described earlier in this
paper. This leads to a large search space for the optimal routine. In
Section~\ref{sec:search-space}, we describe an initial exploration of the
search space of instantiations of sparse matrix times $k$ vectors
multiplication. We conclude this case study with an extensive experimental
evaluation of three sparse algebra kernels, comparing the performance of
automatically generated routines to that of three existing sparse algebra
libraries. We show that significant reductions of execution time compared to
existing library implementations can be achieved and that by relying on a
single library implementation performance is never optimal.

\subsection{Expressing Sparse BLAS routines in Forelem}
\label{sec:forelem-blas}
First, we will demonstrate how Sparse BLAS routines are expressed
in the \emph{forelem} framework\footnote{In a forthcoming
paper we describe an \emph{automated} transformation process that
translates conventionally specified computations into the \emph{forelem}
framework.}.
The expression of a computation in terms of
tuples is the first step in the code and data structure instantiation
process. These tuples arise from ``disassembling'' the original data
structure. A sparse matrix is represented using token tuples of the form
$\langle \mathrm{row, column}\rangle$
with the values of the matrix stored separately as data
tuples. All non-zero matrix elements are extracted from the original data
structure and are represented as token and data tuples. Sparse vectors can
be represented using $\langle \mathrm{index}\rangle$ token tuples and
$\langle \mathrm{value}\rangle$ data
tuples. We consider tuple reservoirs to only contain a single tuple for
every unique $\langle \mathrm{row, column}\rangle$ pair or
$\langle \mathrm{index}\rangle$.

\begin{figure}
\small
\begin{alltt}
\textbf{for} (i = 1; i <= N; i++)
\{
  int sum = 0;
  \textbf{forelem} (t; t \(\in\) T.row[i])
    sum += B[t.col] * A[t];
  C[i] = sum;
\}
\end{alltt}%
\normalsize%
\rule{\linewidth}{0.25pt}%
\caption{Sparse Matrix times Vector Multiplication.}
\label{fig:forelem-mvm-dense}
\end{figure}

As a first routine, we consider the Matrix-Vector Multiplication $C = AB$.
Figure~\ref{fig:forelem-mvm-dense} shows the \emph{forelem} representation
of this multiplication. The token tuples corresponding to the sparse matrix
are stored in tuple reservoir \verb!T!, whereas the values of the matrix
elements are stored in $A$. $C$ and $B$ are considered to be dense vectors
(but do not necessarily have to be dense). A condition on the tuple
reservoir is used in the \emph{forelem} loop to define which tuples should
be processed within an iteration of the outer loop.

\begin{figure}
\small
\begin{alltt}
\textbf{for} (i = N; i >= 1; i--)
\{
  \textbf{forelem} (t; t \(\in\) T.(col,row)[(i, i)])
    x[i] = b[i] / A[t];
  \textbf{forelem} (t; t \(\in\) T.col[i])
    b[i] = b[t.row] - A[t] * x[i];
\}
\end{alltt}%
\normalsize%
\rule{\linewidth}{0.25pt}%
\caption{An implementation of Triangular Solve $Ax = b$ written in the \emph{forelem}
intermediate representation.}
\label{fig:forelem-ts}
\end{figure}

Other BLAS routines can be similarly expressed.
Figure~\ref{fig:forelem-ts} depicts an implementation of Triangular Solve $Ax = B$
using \emph{forelem} loops to access a matrix $A$. As an
additional example, Figure~\ref{fig:forelem-lu} shows an implementation of
LU Factorization. Note that in the latter case every inner loop over the
same sparse matrix $A$ defines a different set of matrix elements to be
iterated.

\begin{figure}
\small
\begin{alltt}
\textbf{for} (i = 1; i <= N; i++)
\{
  p = diag(i);
  \textbf{forelem} (t; t \(\in\) T.(col,row)[(i, (i, \(\infty\)))])
  \{
    A[t] = A[t] / p:
    \textbf{forelem} (l; l \(\in\) T.(row,col)[(i, (i, \(\infty\)))])
    \{
      fillin = True;
      \textbf{forelem} (k; k \(\in\) T.(col,row)[(l.col, t.row])])
      \{
        A[k] = A[k] - A[t] * A[l];
        fillin = False;
      \}
      \textbf{if} (fillin)
      \{
        T = T \(\cup\) (k.row, k.col);
        A[k] = -A[t] * A[l];
      \}
    \}
  \}
\}
\end{alltt}%
\normalsize%
\rule{\linewidth}{0.25pt}%
\caption{An implementation of LU Factorization written in the \emph{forelem}
intermediate representation.}
\label{fig:forelem-lu}
\end{figure}

\subsection{Automatic Sparse Data Structure Generation}
\label{sec:sparse-datagen}
We now demonstrate how the techniques of Orthogonalization and
Materialization that were described in Section~\ref{sec:transformations}
are used to instantiate
efficient sparse algebra routines and reassemble the original sparse matrix
data storage automatically. Using these techniques, many different forms of
a sparse algebra routine can be generated, along with different reassemblies
of the original sparse data storage.

\subsubsection{Concretization}
\label{sec:concretization}
The basis for automatic data structure generation consists of an one-to-one
mapping of materialized loop structures and corresponding symbolic \verb!PA!
array onto a physically allocated array. For instance, a symbolic \verb!PA!
array is mapped onto a two-dimensional C array. As a result the
\emph{forelem} loops accessing the \verb!PA! array are also translated to C
\emph{for} loops. At this point, a reassembled copy of the sparse data
structure is instantiated. This process is referred to as Concretization. To
describe the basic concretization transformation, consider the following
materialized loop as an example:

\small
\begin{alltt}
\textbf{forelem} (i; i \(\in \mathbb{N}*\))
  ... PC[i] ...
\end{alltt}
\normalsize

\noindent
As a first step, $\mathbb{N}*$ materialization is applied, resulting in:

\small
\begin{alltt}
\textbf{forelem} (i; i \(\in\) PA\_len)
  ... PC[i] ...
\end{alltt}
\normalsize

\noindent
then the loop can be subsequently concretized to:

\small
\begin{alltt}
\textbf{for} (i = 0; i < PA\_len; i++)
  ... PC[i] ...
\end{alltt}
\normalsize

\noindent
For this particular, single-dimensional case it is possible to map the
symbolic array \verb!PC! to a physically allocated single-dimensional C
array \verb!PC!.  Note that this storage format is the result of just merely
a straightforward mapping of the materialized sequence into a
(multi)dimensional data structure. This cannot be compared to substituting
coordinate storage by jagged diagonal storage, or the immediate selection of
such a pre-defined format.

\subsubsection{Derivation of Different Data Structures}
\begin{figure}
\begin{center}
\includegraphics[scale=0.5]{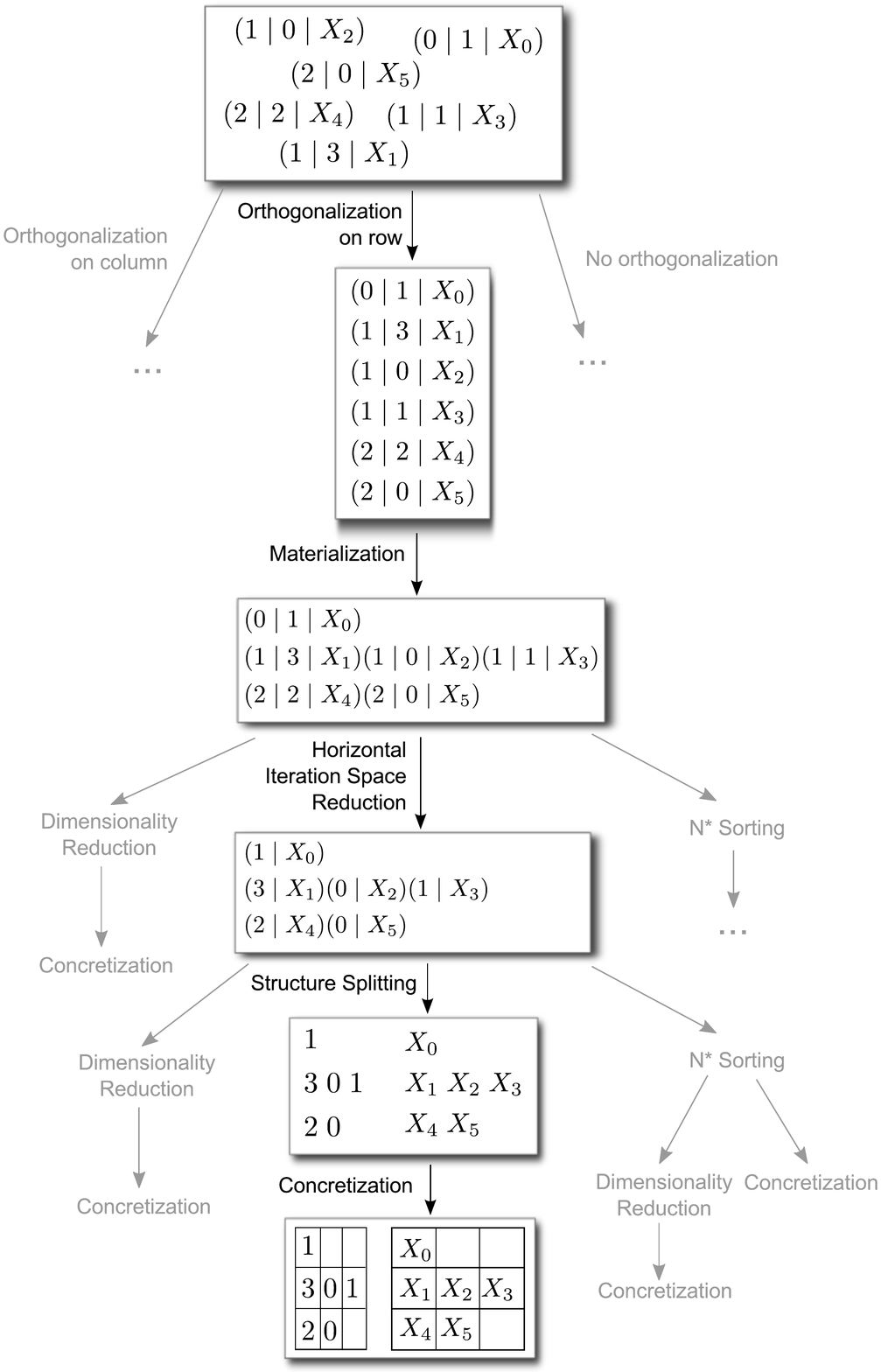}
\caption{An illustration of the application of orthogonalization,
materialization and concretization on sparse matrix tuples in
$(\mathtt{row}\mid\mathtt{col}\mid\mathtt{value})$ format (token and
data tuples combined for conciseness). The
result of this concretization is commonly known as the ITPACK format
(assuming the arrays are stored in column-major order). The arrows
displayed in gray depict a non-exhaustive set of other possibilities.}
\label{fig:storage-formats}
\end{center}
\end{figure}

Figure~\ref{fig:storage-formats} describes the whole process of generating a
data storage format, starting from an unordered set of tuples and using
several transformations of Section~\ref{sec:transformations}.  Note that the
resulting generated storage scheme is described in the literature as
simplified Jagged Diagonal Storage, or ITPACK storage~\cite{bai-2000}.
Important is that this storage format has been deduced without any
predefinition of this format in the \emph{forelem} framework. Rather, it has
been derived with the described sequence of transformations.

In the figure, other transformations that can be applied are shown in gray.
For example, when the structure splitting transformation in the figure is
followed by dimensionality reduction, Compressed Row Storage (CSR) format is
generated. Alternatively, a transformation sequence that continues from
orthogonalization on column can result in Compressed Column Storage (CCS)
format.

Let us consider a different application of the transformations.  As a
starting point, consider the data structure after application of Structure
Splitting in the figure. First, $\mathbb{N}*$ sorting is applied, so that
rows with similar number of entries (non-zeros) are placed close to each
other. This can be helpful as a form of load balancing in the case of
parallelization. Although this transformation changes the order in which the
rows are processed, this does not introduce a problem because before the
concretization the iteration of the rows is specified as a \emph{forelem}
loop which does not impose a particular execution order.

Secondly, the two-dimensional storage structure is put in column-major
order. This can be accomplished by using the Loop Interchange
transformation, as has been described earlier in this section. Thirdly, the
alternative form of $\mathbb{N}*$ materialization is applied. As a result, a
set \verb!PA_len! is generated such that no zeros have to be inserted into
the data structure as padding. Through the application of dimensionality
reduction, the rows of the two-dimensional storage (which thus store columns
of the matrix) are stored back to back in a vector, without insertion of
padding zeroes. As a consequence, an additional data structure is added to
record the start of each row of the original two-dimensional structure.
This concretization leads to the Jagged Diagonal Storage format
described in the literature~\cite{bai-2000}.

Note again that all data structures that come forth out of this automatic
transformation process are basically the result of the application of some
very basic transformations. For instance, ITPACK, Jagged Diagonal Storage
and ELLPACK have been introduced in the literature as smart algorithmic
solutions to optimize sparse matrix times vector computations. As a result
of the transformations described in the preceding sections we can conclude
that these supposedly smart implementations are nothing more than a
succession of very simple basic transformations.

\subsubsection{Generation of Hybrid Storage Formats}
The derivations of sparse data storage formats that have been discussed so
far do not include the loop blocking transformation. Similar to the
well-known loop blocking compiler optimization transformation, loop blocking
can also be applied to \emph{forelem} loops resulting in a nested
\emph{forelem} loop in which the innermost loop iterates over a ``block'' of
the sparse matrix. Then, by including the loop blocking
transformation in the transformation chain, hybrid storage
formats can be generated automatically. Given a loop nest that is
orthogonalized for either the row or column direction, loop blocking
will result in the matrix being processed in blocks of row or column
vectors. When materialization is applied, for each of these blocks a
different set of transformations could be carried out, leading to different
storage formats. So, different hybrid storage formats can be generated by
instantiating different combinations of the basic data structures that can
be generated without loop blocking.

Further possibilities exist if a loop nest is orthogonalized for both the
row and column directions. In this case, two additional outer loops are
added during orthogonalization that control row and column iteration
respectively, as can been seen in Figure~\ref{fig:loop-blocking} (top).
When loop blocking is applied in this case, the matrix is processed by
considering submatrices (blocks) of the matrix. For each submatrix, a
storage format can be instantiated.  The code fragment at the bottom of this
figure shows the result of the application of the loop blocking
transformation.

During Materialization, the three inner loops of this loop nest will be
affected. This code will be transformed such that for each block (submatrix)
that is iterated by the outer loops, a different code sequence carrying out
the computation can be defined that operates on a different automatically
generated data structure. The different data structures that are generated
for these submatrices together form the hybrid data structure for the full
matrix.

\begin{figure}
\small
\begin{alltt}
\textbf{forelem} (i; i \(\in \mathbb{N}\sb{n}\))
  \textbf{forelem} (j; j \(\in \mathbb{N}\sb{m}\))
    \textbf{forelem} (t; t \(\in\) T.(row,col)[(i,j)])
      C[i] += B[t.col] * A[t];
\end{alltt}
\begin{alltt}
\textbf{forelem} (ii; ii \(\in \mathbb{N}\sb{n/x}\))
  \textbf{forelem} (jj; jj \(\in \mathbb{N}\sb{m/y}\))
    \textbf{forelem} (i; i \(\in \mathbb{N}\sb{[ii{\cdot}x, (ii+1){\cdot}x)}\))
      \textbf{forelem} (j; j \(\in \mathbb{N}\sb{[jj{\cdot}y, (jj+1){\cdot}y)}\))
        \textbf{forelem} (t; t \(\in\) T.(row,col)[(i,j)])
          C[i] += B[t.col] * A[t];
\end{alltt}%
\normalsize%
\rule{\linewidth}{0.25pt}%
\caption{Top: Sparse-Matrix Vector Multiplication wherein both row and
column iteration are orthogonalized. Bottom: Resulting of the application of
loop blocking.}%
\label{fig:loop-blocking}
\end{figure}

\subsubsection{Automatic Data Partitioning and Distribution}
In addition to the transformations that have been described in this paper,
further advanced optimizations can be implemented within the \emph{forelem}
framework with ease. This includes transformations that can enable
the framework to generate (distributed and parallel) data structures for
multi-core and distributed execution of sparse matrix algebra. For
instance, for parallel execution a sparse matrix is to be partitioned.
This partitioning can be accomplished with the loop blocking transformation.
Different methods for the partitioning and distribution of the matrix data
can be implemented simply by introducing different methods for the
partitioning of the iteration domain.

For example, Vastenhouw and Bisseling~\cite{vastenhouw-2005} describe a
two-dimensional data distribution method for parallel sparse matrix-vector
multiplication. This data distribution is derived by partitioning the
nonzeros of the sparse matrix such that blocks arise with approximately
similar amounts of nonzeros. In Figure~\ref{fig:loop-blocking}, we showed a
loop structure that gives rise to two-dimensional blocks of data. We defined
these blocks to be regular in size. However, by simply redefinining the
partitioning of $\mathbb{N}_m$ and $\mathbb{N}_n$, we can automatically
generate data structures and data distributions in which the matrix is
partitioned into irregularly sized blocks, similar to that of Vastenhouw and
Bisseling.

Many other methodologies for the distribution of data have been described in
the literature that partition a matrix for efficient distributed processing
or for more efficient use of the
cache~\cite{catalyurek-1999,yzelman-2009,vuduc-2005,pinar-1999}. In fact,
all these distributions are the direct result of the application of the
transformations described in this paper. Further, if we would generalize the
orthogonalization transformation such that arbitrary subsets are selected in
turn, then we would also be able to automatically generate complex, hybrid
data structures, such as described in~\cite{karakasis-2013}.

\subsection{The Transformation Search Space}
\label{sec:search-space}
Using the approach described in the previous section, many different
instantiations of the same sparse matrix routine can be generated. These
different instantiations can be distinguished by the transformations that
were performed before and/or after Materialization, and the order in which
the transformations were performed. In fact, the myriad ways these
transformations can be composed, and can be further combined with parametric
compiler optimizations such as loop unrolling and loop blocking, yields a
large search space. Getting to the code variant with optimal performance for
a particular routine is the problem of finding a particular sequence of code
transformations, or phases. This problem is known in the literature as the
phase-ordering problem~\cite{touati-2006}. While this paper does not intend
to solve the phase-ordering problem of which transformations to apply, we do
discuss an initial exploration and characterization of this complicated
search space, as it is extremely important to understand the characteristics
of this search space before applying techniques to address the phase-ordering
problem. In particular, this section section presents the results of the
initial exploration of the search space of instantiations of sparse matrix
times $k$ vectors multiplication.

\begin{figure}[t]
\begin{center}
\includegraphics[scale=0.30]{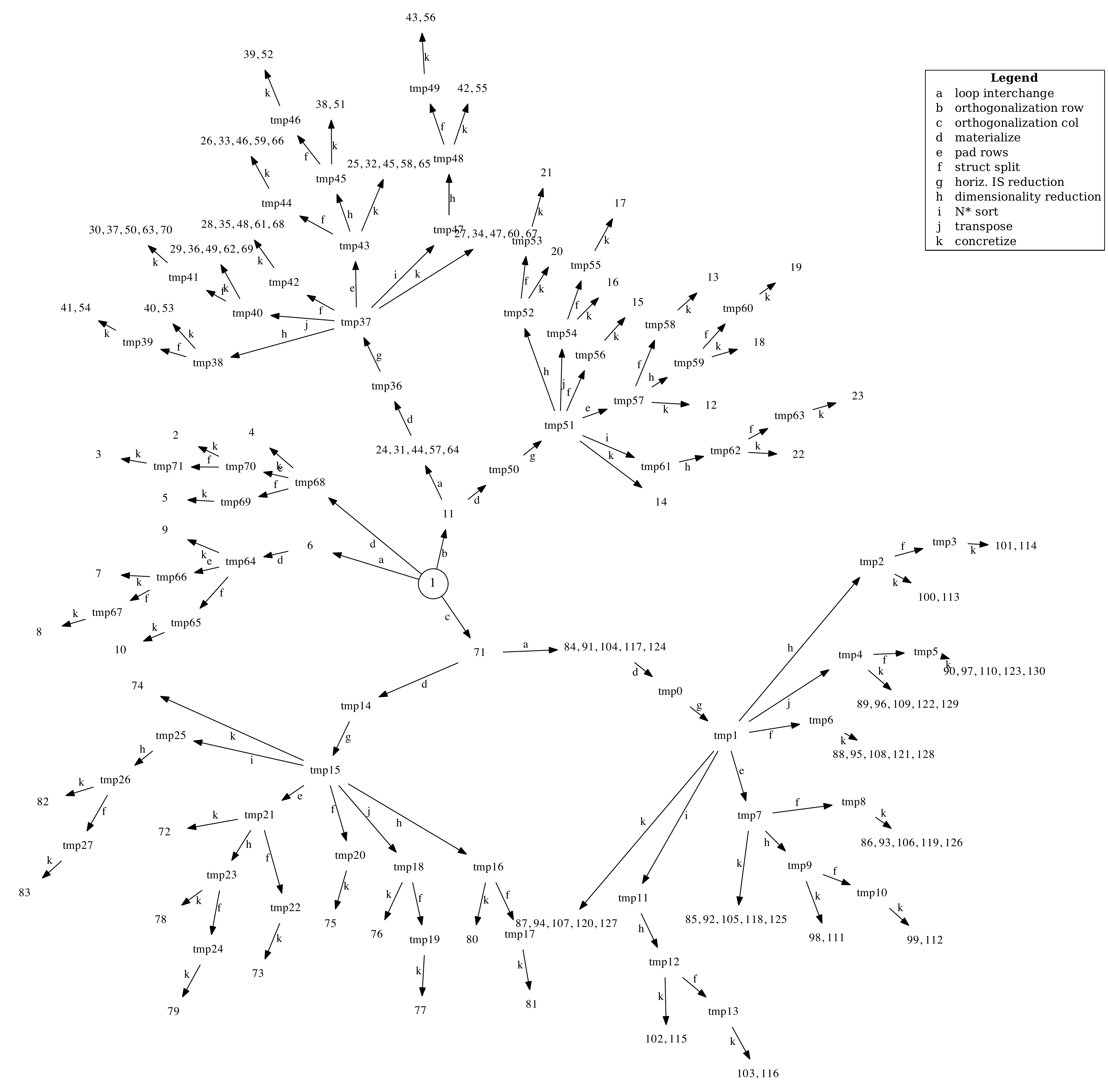}
\caption{The full transformation tree of sparse matrix times $k$ vector
multiplication.}
\label{fig:tree-full}
\end{center}
\end{figure}

The full transformation tree of sparse matrix times $k$ vector
multiplication is shown in
Figure~\ref{fig:tree-full}.
The
starting point of the transformation space is labeled with \emph{1} and
shown in the center of the figure. This is the minimal representation of the
computation as a \emph{forelem} loop. From this point, there are several
different branches of transformations as shown in the picture, resulting in
many different variants, or principal forms. Whenever the label of a node is
prefixed with ``tmp'', the node represents a stage for which no executable
is generated. In all other cases, the executables (variants) are labeled
from 1 to 130. Next to these 130 different implementations, also 25
different data structures are generated, ranging from simple coordinate
storage to compressed row or column schemes, with or without zero-padded
rows or columns, and jagged diagonal like schemes wherein the rows of the
matrix have been permuted or not.
This exemplifies the strength of our approach when compared to sparse
algebra libraries, that on average implement 4, or less, pre-defined sparse
data storage formats. Moreover, when combined with parametric compiler
optimizations such as loop unrolling and loop blocking, the number of
possible code variants is enlarged even further. In the next sections, we
show that in this search space an automatically instantiated routine and
corresponding data structure can be found that outperforms the
implementations of existing sparse algebra libraries.

Finally, note that in this transformation tree we did not consider loop
blocking. If loop blocking would have been taking into account, a multitude
of different hybrid storage formats would have been generated, as has been
illustrated in the previous section. For all these data structures, also
corresponding initialization procedures are automatically generated.

\subsection{Experimental Evaluation}
\label{sec:evaluation}
To evaluate the effectiveness of the framework proposed in this paper, we
compare the performance of the automatically generated variants to routines
from three existing sparse algebra libraries. For three different sparse
algebra kernels, we have generated all possible variants and executed these
for 20 different sparse matrices. Subsequently, we have performed a number
of different analysis to survey the best performing routines, the varying
performance of library routines and to quantity overall library performance.
The results show that variants generated with the framework proposed in this
paper achieve significant reductions of the execution time compared to
existing library implementations by up to 80.8\%. Furthermore, we have found
that none of the surveyed libraries performs consistently better than other
libraries and that relying on a single library implementation never results
in optimal performance across different input matrices. This reinforces the
benefits of our approach which can automatically generate library routines
specific for particular input matrices.

\subsubsection{Experimental Setup}
For the comparison of the automatically generated variants with sparse
algebra libraries, three libraries and their associated data structures
were used:

\begin{itemize}[noitemsep]
\item Blaze 1.2~\cite{iglberger-2012}, with the matrix stored in both
row-major and column-major order;
\item MTL4~\cite{gottschling-2007}, with the
matrix stored in both row-major and column-major order;
\item SparseLib++ 1.7~\cite{dongarra-1994}, with the matrix stored in coordinate storage
format, compressed row storage format and compressed column storage format.
\end{itemize}

\noindent
So, in total 7 different variants from 3 different libraries have been used
in the experimentation.
From these library routines three (shared) computational kernels were
chosen:

\begin{itemize}[noitemsep]
\item sparse matrix times vector multiplication,
\item sparse matrix times matrix multiplication (with a $100$-column dense
matrix),
\item lower triangular solve with unit matrices.
\end{itemize}

\noindent
The comparison with sparse matrix matrix multiplication
from existing sparse algebra libraries has only been carried out with Blaze
and MTL4, because SparseLib++ did not contain API for this computation.
Similarly, the comparison with triangular solve was only done with the MTL4
and SparseLib++ libraries.

The experiments have been performed on two architectures. The first
architecture consists of an
Intel Xeon 5150 CPU at 2.66 Ghz, with 16GB RAM, running Ubuntu Linux
10.04.4. The second architecture is a machine that consists of an Intel
Xeon E5-2650 CPU at 2.00 GHz, with 64 GB RAM, running CentOS 5.0. The
compiler used in both cases is gcc 4.4.  These architectures will be
referred to as the Xeon 5150 and Xeon E5 architectures respectively.  Twenty
matrices have been surveyed, taken from the University of Florida Matrix
Collection~\cite{davis-matrix}. To remove fluctuation from the results, the
computation performed by each variant or library is repeated 10 times.

%
%

\begin{table}
\small
\centering
\begin{tabular}{|l|c|c|c|c|c|c|c|c|c|}
\hline
\multicolumn{8}{|c|}{\textbf{\normalsize{(a) Xeon 5150}}} \\
\hline
 & \multicolumn{2}{c|}{\textbf{Blaze}} &
\multicolumn{2}{c|}{\textbf{MTL4}} &
\multicolumn{3}{c|}{\textbf{SL++}} \\
 & \textbf{CRS} & \textbf{CCS} &
\textbf{CRS} & \textbf{CCS} &
\textbf{COO} & \textbf{CRS} & \textbf{CCS} \\
\hline
\begin{filecontents*}{csvdata/spmv-hpc0-speedup.csv}
Matrix,Maximum,Minimum,Blaze CRS,Blaze CCS,MTL4 CRS,MTL4 CCS,SPARSE1.3,SL++ COO,SL++ CRS,SL++ CCS
Erdos971,77.7\%,37.1\%,52.7\%,61.5\%,63.1\%,\maxperc{77.4\%},77.7\%,\minperc{37.1\%},59.0\%,58.4\%
mcfe,78.5\%,24.2\%,25.7\%,54.5\%,48.7\%,\maxperc{59.5\%},78.5\%,36.1\%,\minperc{24.2\%},32.4\%
blckhole,70.6\%,20.9\%,\minperc{20.9\%},47.5\%,48.0\%,\maxperc{63.6\%},70.6\%,43.5\%,34.0\%,41.1\%
c-62,94.7\%,24.8\%,33.9\%,41.3\%,29.9\%,45.7\%,94.7\%,\maxperc{61.4\%},\minperc{24.8\%},34.1\%
OPF\_10000,92.5\%,31.7\%,45.0\%,39.9\%,\minperc{31.7\%},45.5\%,92.5\%,\maxperc{64.7\%},36.2\%,40.4\%
lhr71,89.9\%,18.8\%,34.0\%,41.3\%,30.6\%,\maxperc{42.9\%},{89.9\%},31.2\%,24.0\%,\minperc{18.8\%}
stomach,89.8\%,24.2\%,35.4\%,\maxperc{54.6\%},29.8\%,38.6\%,{89.8\%},37.9\%,\minperc{24.2\%},28.0\%
Orsreg\_1,69.7\%,21.2\%,\minperc{21.2\%},45.8\%,47.4\%,\maxperc{62.8\%},{69.7\%},31.8\%,25.7\%,34.2\%
shipsec1,93.2\%,8.2\%,32.9\%,42.6\%,29.4\%,48.5\%,{93.2\%},\maxperc{61.5\%},\minperc{8.2\%},21.5\%
shipsec5,93.1\%,8.5\%,33.0\%,42.9\%,29.7\%,49.2\%,{93.1\%},\maxperc{61.8\%},\minperc{8.5\%},21.8\%
pdb1HYS,92.8\%,0.0\%,30.5\%,37.6\%,26.6\%,41.4\%,{92.8\%},\maxperc{60.2\%},\minperc{0.0\%},9.3\%
or2010,83.5\%,25.3\%,28.1\%,27.3\%,\minperc{25.3\%},44.5\%,{83.5\%},\maxperc{45.9\%},33.2\%,34.4\%
Para-4,92.1\%,10.0\%,27.9\%,37.7\%,26.9\%,\maxperc{38.7\%},{92.1\%},27.4\%,\minperc{10.0\%},21.2\%
G2\_circuit,89.5\%,30.9\%,34.4\%,39.6\%,\minperc{30.9\%},38.9\%,{89.5\%},\maxperc{63.4\%},52.8\%,53.1\%
144,91.4\%,17.7\%,30.5\%,37.0\%,29.7\%,36.6\%,{91.4\%},\maxperc{47.8\%},\minperc{17.7\%},20.9\%
cop20k\_A,92.3\%,16.1\%,31.5\%,39.8\%,27.7\%,37.8\%,{92.3\%},\maxperc{57.9\%},\minperc{16.1\%},22.8\%
consph,93.2\%,8.9\%,33.7\%,42.6\%,30.6\%,48.0\%,{93.2\%},\maxperc{62.1\%},\minperc{8.9\%},23.2\%
Raj1,88.8\%,29.9\%,29.9\%,35.5\%,\minperc{29.9\%},37.7\%,{88.8\%},\maxperc{45.6\%},42.6\%,42.6\%
3dtube,94.3\%,6.3\%,33.8\%,42.9\%,30.2\%,48.8\%,{94.3\%},\maxperc{62.0\%},\minperc{6.3\%},21.2\%
net150,94.1\%,18.5\%,31.6\%,39.4\%,38.0\%,49.0\%,{94.1\%},\maxperc{61.2\%},\minperc{18.5\%},25.4\%
\end{filecontents*}
\csvreader[late after line=\\]{csvdata/spmv-hpc0-speedup.csv}{}%
{\csvcoli & \csvcoliv & \csvcolv & \csvcolvi & \csvcolvii & \csvcolix & \csvcolx & \csvcolxi}
\hline
\hline
\multicolumn{8}{|c|}{\textbf{\normalsize{(b) Xeon E5}}} \\
\hline
 & \multicolumn{2}{c|}{\textbf{Blaze}} &
\multicolumn{2}{c|}{\textbf{MTL4}} &
\multicolumn{3}{c|}{\textbf{SL++}} \\
 & \textbf{CRS} & \textbf{CCS} &
\textbf{CRS} & \textbf{CCS} &
\textbf{COO} & \textbf{CRS} & \textbf{CCS} \\
\hline
\begin{filecontents*}{csvdata/spmv-LZ-speedup.csv}
Matrix,Maximum,Minimum,Blaze CRS,Blaze CCS,MTL4 CRS,MTL4 CCS,SPARSE1.3,SL++ COO,SL++ CRS,SL++ CCS
Erdos971,71.6\%,18.0\%,\minperc{18.0\%},52.4\%,40.7\%,\maxperc{70.7\%},{71.6\%},29.5\%,38.6\%,47.9\%
mcfe,87.2\%,7.4\%,\minperc{7.4\%},49.3\%,31.1\%,\maxperc{59.3\%},{87.2\%},30.7\%,20.1\%,30.9\%
blckhole,69.9\%,9.4\%,\minperc{9.4\%},45.4\%,31.8\%,\maxperc{65.8\%},{69.9\%},42.6\%,35.2\%,43.3\%
c-62,94.8\%,7.2\%,\minperc{7.2\%},45.0\%,24.3\%,\maxperc{55.3\%},{94.8\%},44.1\%,29.7\%,47.9\%
OPF\_10000,88.3\%,7.4\%,\minperc{7.4\%},47.3\%,28.5\%,\maxperc{56.9\%},{88.3\%},44.4\%,37.4\%,48.0\%
lhr71,93.1\%,22.4\%,25.7\%,48.5\%,31.7\%,\maxperc{57.7\%},{93.1\%},34.5\%,\minperc{22.4\%},36.4\%
stomach,92.1\%,20.1\%,\minperc{20.1\%},49.8\%,23.0\%,\maxperc{54.4\%},{92.1\%},37.1\%,30.2\%,43.2\%
Orsreg\_1,73.1\%,10.4\%,\minperc{10.4\%},44.3\%,29.8\%,\maxperc{65.1\%},{73.1\%},29.0\%,27.2\%,36.7\%
shipsec1,95.4\%,17.8\%,30.3\%,\maxperc{52.9\%},27.1\%,49.0\%,{95.4\%},41.1\%,\minperc{17.8\%},46.3\%
shipsec5,95.3\%,17.8\%,31.7\%,\maxperc{53.1\%},28.4\%,49.7\%,{95.3\%},39.6\%,\minperc{17.8\%},46.3\%
pdb1HYS,96.5\%,5.1\%,11.4\%,45.1\%,11.5\%,\maxperc{45.5\%},{96.5\%},34.6\%,\minperc{5.1\%},25.1\%
or2010,91.8\%,46.6\%,\minperc{46.6\%},51.3\%,48.9\%,\maxperc{69.2\%},{91.8\%},47.7\%,51.9\%,55.2\%
Para-4,95.8\%,18.0\%,31.4\%,\maxperc{48.1\%},\minperc{18.0\%},46.0\%,{95.8\%},27.3\%,21.6\%,47.6\%
G2\_circuit,83.2\%,20.9\%,\minperc{20.9\%},54.1\%,30.7\%,\maxperc{64.6\%},{83.2\%},58.2\%,54.5\%,59.3\%
144,93.0\%,12.9\%,\minperc{12.9\%},35.8\%,21.5\%,\maxperc{43.7\%},{93.0\%},29.7\%,14.0\%,25.2\%
cop20k\_A,94.6\%,17.9\%,19.3\%,\maxperc{48.4\%},19.8\%,43.2\%,{94.6\%},35.5\%,\minperc{17.9\%},40.0\%
consph,96.9\%,15.1\%,18.7\%,\maxperc{53.3\%},19.3\%,48.9\%,{96.9\%},40.8\%,\minperc{15.1\%},36.6\%
Raj1,89.3\%,22.8\%,\minperc{22.8\%},48.6\%,29.3\%,\maxperc{59.3\%},{89.3\%},44.5\%,43.5\%,48.9\%
3dtube,96.8\%,17.9\%,27.5\%,\maxperc{54.1\%},27.0\%,50.5\%,{96.8\%},41.8\%,\minperc{17.9\%},40.8\%
net150,96.7\%,13.1\%,17.2\%,\maxperc{49.1\%},18.4\%,48.1\%,{96.7\%},40.2\%,\minperc{13.1\%},34.0\%
\end{filecontents*}
\csvreader[late after line=\\]{csvdata/spmv-LZ-speedup.csv}{}%
{\csvcoli & \csvcoliv & \csvcolv & \csvcolvi & \csvcolvii & \csvcolix & \csvcolx & \csvcolxi}
\hline
\end{tabular}
\normalsize
\caption{Overview of the reduction of the execution time of
\emph{sparse matrix times vector multiplication} achieved
by variants generated by our approach compared to library implementations.}
\label{tab:spmv-speedup}
\end{table}

\subsubsection{Results}
Experiments have been conducted with each of the above mentioned
computational kernels. Each of these kernels has been run for each of the
twenty matrices on both architectures. Note that our method relies on the
fact that for each combination of matrix, kernel and architecture an
optimized executable is automatically generated. The potential for
performance optimization is demonstrated in this section by comparing each
library routine with the best performing automatically generated kernel.
The percentage reduction of the single-core execution time of the
automatically generated variant over the library routines has been computed
and is shown in the tables. In these tables, results printed on a
black background indicate the largest reduction achieved by an automatically
generated variant over a library routine. The results printed on a gray
background indicate the minimum performance improvement that is realized,
i.e., the automatically generated kernel is compared to the best performing
routine with the best data structure selection.

Table~\ref{tab:spmv-speedup}(a) shows the comparison of the sparse matrix
times vector multiplication kernel on the Xeon 5150 architecture.
Improvements are achieved over all library kernels. The achieved reductions
go up to 77.4\% for the \emph{Erdos971} matrix using the MTL4 CCS routine.
This means that the automatically generated variant runs in 22.6\% of the
execution time of the MTL4 CCS routine.

Instead of considering the library routines individually, the collection of
library routines can be considered as a whole. Even if we take the fastest
library routine for each matrix in this case, our method still achieves
performance reductions of 37.1\% (\emph{Erdos971}), 31.7\%
(\emph{OPF\_10000}) and 30.9\% (\emph{G2\_circuit}) for the Xeon 5150
architecture.

In Table~\ref{tab:spmv-speedup}(b) the results of the same kernel are shown
for the Xeon E5 architecture. From the table can be seen that reductions
of up to 70.7\% (\emph{Erdos971} matrix, MTL4 CCS routine) are achieved. The
performance reduction compared to the fastest library routine per matrix
is still up to 46.6\% (\emph{or2010}).

\begin{table}
\centering
\small
\begin{tabular}{|l|c|c|c|c||c|c|c|c|}
\hline
 & \multicolumn{4}{c||}{\textbf{\normalsize{Xeon 5150}}} &
\multicolumn{4}{c|}{\textbf{\normalsize{Xeon E5}}} \\
\hline
 & \multicolumn{2}{c|}{\textbf{Blaze}} &
\multicolumn{2}{c||}{\textbf{MTL4}} &
\multicolumn{2}{c|}{\textbf{Blaze}} &
\multicolumn{2}{c|}{\textbf{MTL4}} \\
 &  \textbf{CRS} & \textbf{CCS} & \textbf{CRS} & \textbf{CCS} &
 \textbf{CRS} & \textbf{CCS} & \textbf{CRS} & \textbf{CCS} \\
\hline
\begin{filecontents*}{csvdata/spmm-both-speedup.csv}
matrix,Blaze CRS,Blaze CCS,MTL4 CRS,MTL4 CCS,Blaze CRS,Blaze CCS,MTL4 CRS,MTL4 CCS
Erdos971,\minperc{56.2\%},73.4\%,63.8\%,\maxperc{76.4\%},\minperc{52.4\%},71.5\%,64.2\%,\maxperc{75.7\%}
mcfe,49.0\%,75.0\%,\minperc{38.1\%},\maxperc{78.7\%},44.7\%,74.2\%,\minperc{35.9\%},\maxperc{79.0\%}
blckhole,60.3\%,73.1\%,\minperc{56.8\%},\maxperc{75.9\%},\minperc{60.6\%},74.4\%,62.8\%,\maxperc{78.1\%}
c-62,\minperc{30.6\%},56.4\%,34.8\%,\maxperc{62.0\%},\minperc{39.0\%},67.8\%,44.9\%,\maxperc{73.4\%}
OPF\_10000,\minperc{37.2\%},59.7\%,44.2\%,\maxperc{65.1\%},\minperc{42.1\%},68.7\%,52.0\%,\maxperc{74.6\%}
lhr71,41.3\%,68.4\%,\minperc{30.9\%},\maxperc{72.9\%},44.1\%,72.4\%,\minperc{38.5\%},\maxperc{77.5\%}
stomach,47.0\%,67.4\%,\minperc{37.0\%},\maxperc{71.3\%},49.9\%,71.8\%,\minperc{45.6\%},\maxperc{76.6\%}
Orsreg\_1,58.0\%,71.9\%,\minperc{57.8\%},\maxperc{75.2\%},\minperc{57.3\%},72.6\%,62.0\%,\maxperc{76.4\%}
shipsec1,43.7\%,71.6\%,\minperc{33.2\%},\maxperc{75.6\%},42.6\%,73.6\%,\minperc{29.4\%},\maxperc{78.6\%}
shipsec5,44.1\%,71.9\%,\minperc{32.4\%},\maxperc{75.7\%},43.0\%,73.7\%,\minperc{29.1\%},\maxperc{78.6\%}
pdb1HYS,43.4\%,72.8\%,\minperc{26.4\%},\maxperc{76.9\%},42.1\%,73.9\%,\minperc{23.1\%},\maxperc{79.0\%}
or2010,\minperc{26.5\%},42.4\%,32.8\%,\maxperc{49.9\%},\minperc{40.0\%},61.0\%,57.0\%,\maxperc{67.0\%}
Para-4,\minperc{29.6\%},61.5\%,35.6\%,\maxperc{67.2\%},35.2\%,68.6\%,\minperc{33.3\%},\maxperc{74.4\%}
G2\_circuit,\minperc{35.2\%},55.5\%,42.1\%,\maxperc{59.6\%},\minperc{44.9\%},67.3\%,62.7\%,\maxperc{71.9\%}
144,\minperc{14.6\%},41.7\%,17.7\%,\maxperc{50.1\%},\minperc{12.4\%},48.5\%,31.9\%,\maxperc{60.3\%}
cop20k\_A,\minperc{29.8\%},56.9\%,33.6\%,\maxperc{62.0\%},\minperc{36.5\%},66.9\%,39.6\%,\maxperc{72.7\%}
consph,37.3\%,69.3\%,\minperc{24.1\%},\maxperc{74.0\%},41.4\%,73.6\%,\minperc{26.1\%},\maxperc{78.6\%}
Raj1,\minperc{31.6\%},51.0\%,44.5\%,\maxperc{55.4\%},\minperc{37.5\%},64.7\%,60.8\%,\maxperc{69.0\%}
3dtube,43.3\%,71.7\%,\minperc{32.9\%},\maxperc{75.7\%},41.9\%,73.5\%,\minperc{27.0\%},\maxperc{78.6\%}
net150,\minperc{25.7\%},59.6\%,43.2\%,\maxperc{65.1\%},32.5\%,68.5\%,\minperc{30.9\%},\maxperc{74.7\%}
\end{filecontents*}
\csvreader[late after line=\\]{csvdata/spmm-both-speedup.csv}{}%
{\csvcoli & \csvcolii & \csvcoliii & \csvcoliv & \csvcolv & \csvcolvi &
\csvcolvii & \csvcolviii & \csvcolix }
\hline
\end{tabular}
\normalsize
\caption{Reduction of execution time of \emph{sparse matrix times matrix
multiplication} routines.}
\label{tab:spmm-speedup}
\end{table}

In Table~\ref{tab:spmm-speedup} the percentage reduction of the
single-core execution time of the sparse matrix times matrix kernel is
presented for the libraries that implement this routine. Performance
reductions are achieved of up to 79.0\%. From the numbers on a gray
background in this table can be seen that for both architectures variants
are generated that decrease the execution time beyond 50\% compared to the
best-performing library routine for that matrix. The minimum reductions
that are attained are significant: at least a 24.1\% reduction for the
Xeon 5150 architecture (\emph{consph} matrix) and 12.4\% for the Xeon E5
architecture (\emph{144} matrix).

\begin{table}
\centering
\small
\begin{tabular}{|l|c|c|c|c||c|c|c|c|}
\hline
 & \multicolumn{4}{c||}{\textbf{\normalsize{Xeon 5150}}} &
\multicolumn{4}{c|}{\textbf{\normalsize{Xeon E5}}} \\
\hline
 & \multicolumn{2}{c|}{\textbf{MTL4}} &
\multicolumn{2}{c||}{\textbf{SL++}} &
\multicolumn{2}{c|}{\textbf{MTL4}} &
\multicolumn{2}{c|}{\textbf{SL++}} \\
 &  \textbf{CRS} & \textbf{CCS} & \textbf{CRS} & \textbf{CCS} &
 \textbf{CRS} & \textbf{CCS} & \textbf{CRS} & \textbf{CCS} \\
\hline
\begin{filecontents*}{csvdata/sptrsv-both-speedup.csv}
matrix,MTL4 CRS,MTL4 CCS,SL++ CRS,SL++ CCS,MTL4 CRS,MTL4 CCS,SL++ CRS,SL++ CCS
Erdos971,79.3\%,\maxperc{80.1\%},\minperc{52.6\%},55.9\%,\maxperc{80.8\%},78.6\%,\minperc{26.0\%},43.8\%
mcfe,\maxperc{65.6\%},64.8\%,\minperc{7.9\%},14.6\%,68.7\%,\maxperc{70.2\%},\minperc{3.6\%},21.5\%
blckhole,63.1\%,\maxperc{67.6\%},\minperc{0.9\%},9.4\%,\maxperc{70.8\%},70.1\%,\minperc{0.0\%},23.6\%
c-62,64.5\%,\maxperc{74.9\%},\minperc{10.4\%},22.0\%,66.3\%,\maxperc{71.1\%},\minperc{-4.8\%},15.6\%
OPF\_10000,67.4\%,\maxperc{75.1\%},\minperc{14.6\%},23.1\%,67.6\%,\maxperc{70.4\%},\minperc{-3.4\%},16.5\%
lhr71,51.7\%,\maxperc{63.5\%},17.3\%,\minperc{10.4\%},65.5\%,\maxperc{74.7\%},\minperc{-1.6\%},24.9\%
stomach,46.2\%,\maxperc{62.5\%},\minperc{15.3\%},24.2\%,66.7\%,\maxperc{72.9\%},\minperc{-1.6\%},28.0\%
Orsreg\_1,62.5\%,\maxperc{66.3\%},\minperc{-2.5\%},6.0\%,\maxperc{69.1\%},67.9\%,\minperc{-1.7\%},10.8\%
shipsec1,43.6\%,\maxperc{61.0\%},\minperc{-1.4\%},3.4\%,62.0\%,\maxperc{72.2\%},\minperc{-3.0\%},19.4\%
shipsec5,45.9\%,\maxperc{63.3\%},\minperc{3.4\%},8.9\%,62.4\%,\maxperc{72.6\%},\minperc{-2.5\%},20.9\%
pdb1HYS,51.5\%,\maxperc{69.0\%},\minperc{3.9\%},9.4\%,61.5\%,\maxperc{71.2\%},\minperc{-6.4\%},13.9\%
or2010,61.4\%,\maxperc{69.4\%},\minperc{30.9\%},43.7\%,\maxperc{78.5\%},78.2\%,\minperc{44.7\%},48.7\%
Para-4,42.7\%,\maxperc{59.7\%},\minperc{-1.8\%},6.6\%,60.1\%,\maxperc{69.4\%},\minperc{-3.1\%},17.8\%
G2\_circuit,59.3\%,\maxperc{69.2\%},\minperc{31.9\%},38.3\%,62.9\%,\maxperc{64.7\%},\minperc{0.0\%},1.5\%
144,43.2\%,\maxperc{58.4\%},\minperc{1.2\%},14.4\%,60.6\%,\maxperc{67.8\%},\minperc{3.9\%},20.3\%
cop20k\_A,50.0\%,\maxperc{64.4\%},\minperc{8.3\%},16.9\%,62.8\%,\maxperc{70.4\%},\minperc{-1.7\%},18.3\%
consph,42.7\%,\maxperc{62.1\%},\minperc{-2.9\%},2.8\%,61.9\%,\maxperc{72.6\%},\minperc{-0.2\%},19.7\%
Raj1,52.4\%,\maxperc{68.2\%},\minperc{25.7\%},33.6\%,72.8\%,\maxperc{73.6\%},\minperc{10.4\%},23.3\%
3dtube,48.5\%,\maxperc{64.3\%},\minperc{4.2\%},10.1\%,66.1\%,\maxperc{75.3\%},\minperc{2.3\%},23.1\%
net150,48.3\%,\maxperc{66.2\%},\minperc{4.6\%},10.1\%,64.5\%,\maxperc{73.5\%},\minperc{0.4\%},17.5\%
\end{filecontents*}
\csvreader[late after line=\\]{csvdata/sptrsv-both-speedup.csv}{}%
{\csvcoli & \csvcolii & \csvcoliii & \csvcoliv & \csvcolv & \csvcolvi &
\csvcolvii & \csvcolviii & \csvcolix }
\hline
\end{tabular}
\normalsize
\caption{Reduction of execution time of \emph{sparse triangular solve}
routines.}
\label{tab:sptrsv-speedup}
\end{table}

Finally, the results of the experiments with sparse triangular solve are
shown in an identical manner in Table~\ref{tab:sptrsv-speedup}. For the Xeon
5150 architecture automatically generated routines are found that decrease
the execution time by up to 80.1\% and up to 80.8\% for the Xeon E5.  In
most cases, an reduction of 10.4\% to to 31.9\% is achieved compared to the
best-performing library routine. For four cases a slight decrease in
performance is seen although still on par with the fastest library routine
found. In case of the Xeon E5 architecture reductions of the computation
time up to 44.7\% are found. In the latter case also increases in
computation time in almost half of the cases can be found. Note that these
increases are mostly within 2 to 3 percent, so the performance in these
cases can be considered on par with the library routines. The results for
the triangular solve kernel are not as significant as for the other two
computational kernels. This is caused by the fact that for this kernel the
optimization possibilities are very limited because of the absences of data
reuse in combination with data dependencies limiting execution reordering.

\subsubsection{Varying Performance of Library Routines}
From the presented results can also be observed that the performance results
among the libraries vary. There is no library that performs consistently
better than the other libraries for the surveyed kernels, matrices and
architectures. For instance, while MTL4 performs better than Blaze on sparse
matrix times matrix multiplication in half of the cases
(Table~\ref{tab:spmm-speedup}), the same MTL4 library never outperforms SL++
on sparse triangular solve (Table~\ref{tab:sptrsv-speedup}). When
considering the results for sparse matrix times vector multiplication in
Table~\ref{tab:spmv-speedup}, Blaze outperforms other libraries on the Xeon
E5 half of the time, while Blaze only achieves this twice on the Xeon 5150
architecture.

Also among the different data structures implemented by the libraries there
is no data structure that is always better than others. As can be seen in
Table~\ref{tab:spmv-speedup}(a) all data formats, CRS, CCS and coordinate
storage, outperform all other data formats at least once.  In
Table~\ref{tab:sptrsv-speedup} we observe that while SL++ CRS typically
outperforms SL++ CCS, there is one case where SL++ CCS is faster. Also for
the MTL4 library can be seen in the same table that the CCS data structure
sometimes outperforms its CRS data structure.

Furthermore, if a library routine has the worst performance for a number of
matrices, this does not mean that such a routine is never the best performer
among library routines. In fact, whereas the SL++ COO routine is regularly
the most distanced from the performance achieved by our generated variants,
it is also in one case the least distanced.

The extent by which the performance of a library varies among different
kernels, matrices and architectures is significant. For instance, from
Table~\ref{tab:spmv-speedup}(a) can be seen that the generated variant for
the \emph{blckhole} matrix reduces the execution time by 20.9\% compared to
the Blaze CRS kernel. This is the minimum reduction that is achieved by our
method, the realized reductions are larger compared to all other evaluated
libraries. When the reduction of the execution time of just the Blaze CRS is
considered over different matrices, then the measured performance
improvements achieved by our method over this library routine go up to
52.7\%. Note that for the several other matrices these achieved execution
time reductions are smaller, for instance SL++ CRS outperforms the Blaze
routines several times. Even compared between just these two library
routines, the execution time of the Blaze routine can be reduced by up to
30.5\% by the SL++ CRS routine in certain cases. This shows that when
relying on a single library implementation, optimal performance is never
achieved.

\subsubsection{Quantifying Overall Performance Of Library Routines}
The above examples make apparent that the performance of a particular
routine varies when run for different matrices and on different
architectures. Routines that are good performers for certain matrices, can
be among the worst performers for other matrices.  Within the approach
proposed in this paper, many different variants are generated and these
variants all exhibit different performance characteristics.  As a
consequence, in order to demonstrate the many advantages of this framework
it is not sufficient to simply present the speedup of the best performing
variant over an existing routine using an existing data structure.  A
variant that performs well for a certain matrix and architecture might be a
bad performer for another combination.

In order to quantify these differences we use a \emph{coverage} metric which
essentially captures the percentage of matrices for which a single good
performing routine can be found, given a certain percentage such a routine
may be distanced from the optimal implementation.  For example, if the
coverage is less than 60\%, then the found routine does not belong to the
best performing routines for over 40\% of the remaining matrices and
consequently a different routine must be chosen for these matrices in order
to obtain optimal performance. As another example, consider that a routine
has been found with a 100\% coverage and that routines may have been
distanced up to 45\% from the optimal implementation.  Although a single
routine has been found that is among the best performing for all matrices,
this comes at the cost of performing 45\% short of the optimal performance
for certain matrices.

We now formally define this notion of coverage. Let $R$ be a set of
routines, that can be library routines as well as variants generated using
the framework described in this paper. $M$ is the set of all matrices
surveyed in the experimentation. The measured execution time for a routine
$r \in R$ and a matrix $m \in M$ is denoted $\mathrm{exec}(r, m)$. For a
given matrix $m \in M$, one of the routines in $R$ will have the best
execution time\footnote{Note that it is possible for the best execution time
to be shared with more than 1 routine.}, this routine is referred to as $b$.
Now, a \emph{top group} $T \subset R$ can be defined for a matrix $m$ that
contains variants with a runtime at most $t\%$ away from the best variant
$b$:
\begin{equation*}
T(m) = \{ r \in R \mid \mathrm{exec}(r, m) \leq t\% \times \mathrm{exec}(b, m) \}
\end{equation*}

\noindent
Subsequently, the \emph{weight} of a routine $r \in R$ for a collection
of matrices $M$ is defined that indicates the number of matrices of
that collection for which the routine is a good performer:
\begin{equation*}
\mathrm{weight}(r) = | \{m \in M \mid r \in T(m) \} |
\end{equation*}

\noindent
Finally, the \emph{coverage} is the maximal weight found among the
set of routines for a collection of matrices:
\begin{equation*}
\mathrm{coverage} = \operatorname*{max}\limits_{r \in R}\;\mathrm{weight}(r)
\end{equation*}

\noindent
Summarizing, in order to compute the coverage, a collection of
matrices $M$, set of routines $R$ and a percentage $t$ from the optimum for
which variants are still considered to be among the best performing, are
needed. The coverage indicates the maximal number of matrices
for which a single routine is among the best performing routines.
In other words, for how many matrices acceptable performance can be
achieved with a single routine.

Note the effect of changing these parameters: if the percentage $t$
is increased, a larger set of routines is considered the best
performing variants. As a result, the chances that a routine with
a higher coverage is found are increased. However, as routines are
allowed that are further distanced from the optimal implementation,
overall performance of the routine might be worse! The best scenario
is to find the maximal coverage for a value of $t$ which is as small
as possible.

\begin{table}
\centering
\small
\begin{minipage}{0.508\linewidth}
\begin{tabular}{|l||c|c|c|c|c|}
\hline
 & \multicolumn{5}{c|}{\textbf{\normalsize{Xeon 5150}}} \\
\hline
\textbf{t\%}     & 10\% & 20\% & 30\% & 40\% & 50\% \\
\hline
\begin{filecontents*}{csvdata/hpc0-lib-cov.csv}
a,b,c,d,e,f
SPMV,76\%,83\%,90\%,90\%,100\%
SPMM,76\%,84\%,100\%,100\%,100\%
TrSv,100\%,100\%,100\%,100\%,100\%
\end{filecontents*}
\csvreader[late after line=\\]{csvdata/hpc0-lib-cov.csv}{}%
{\csvcoli & \csvcolii & \csvcoliii & \csvcoliv & \csvcolv & \csvcolvi}
\hline
\end{tabular}
\end{minipage}
\begin{minipage}{0.38\linewidth}
\begin{tabular}{|c|c|c|c|c|}
\hline
\multicolumn{5}{|c|}{\textbf{\normalsize{Xeon E5}}} \\
\hline
10\% & 20\% & 30\% & 40\% & 50\% \\
\hline
\begin{filecontents*}{csvdata/LZ-lib-cov.csv}
a,b,c,d,e,f
SPMV,80\%,97\%,100\%,100\%,100\%
SPMM,71\%,77\%,97\%,100\%,100\%
TrSv,100\%,100\%,100\%,100\%,100\%
\end{filecontents*}
\csvreader[late after line=\\]{csvdata/LZ-lib-cov.csv}{}%
{\csvcolii & \csvcoliii & \csvcoliv & \csvcolv & \csvcolvi}
\hline
\end{tabular}\end{minipage}
\normalsize%
\caption{Computed coverages for the collection of library routines for
increasing values of $t\%$.}
\label{tab:lib-t}
\end{table}


Using this metric of coverage the differences in performance between
different library routines are now evaluated. To this end, a collection
has been formed of the kernels from the different libraries and the
coverage was computed for increasing values of $t\%$, see
Table~\ref{tab:lib-t}. From these results can be seen that the minimum value
of $t\%$ that is necessary to find a single best-performing library routine,
that is a library routine with a coverage of 100\%, is significant. For
instance, for sparse matrix times vector multiplication this is 43\% for the
Xeon 5150 and 21\% for the Xeon E5 architectures.  From these results
follows that when relying on a single library implementation of a sparse
matrix kernel to perform computations on different sparse matrices on
different architectures, the optimal performance is never achieved. In fact,
significant performance losses are incurred. The minimum values of $t\%$
also indicate the maximal performance loss that one has to accept in order
to work with a single library routine. In case of the sparse matrix time
vector multiplication, the performance thus falls short by as much as 43\%.

The fact that for triangular solve the values of $t\%$ for a 100\% coverage
are below 100\% correlates with the performance results seen in
Table~\ref{tab:sptrsv-speedup}. For the Xeon E5 architecture $t\% = 0\%$ is
found, which indicates that it is possible to find a single best-performing
library routine that fits all matrices in this case. This further confirms
the fact that the optimization opportunities of the triangular solve kernel
are very limited compared to the other computational kernels.

\subsubsection{Optimization Of Computational Kernels Per Architecture}
\begin{figure}
\begin{center}
\begin{minipage}{0.49\linewidth}
\includegraphics[scale=0.32]{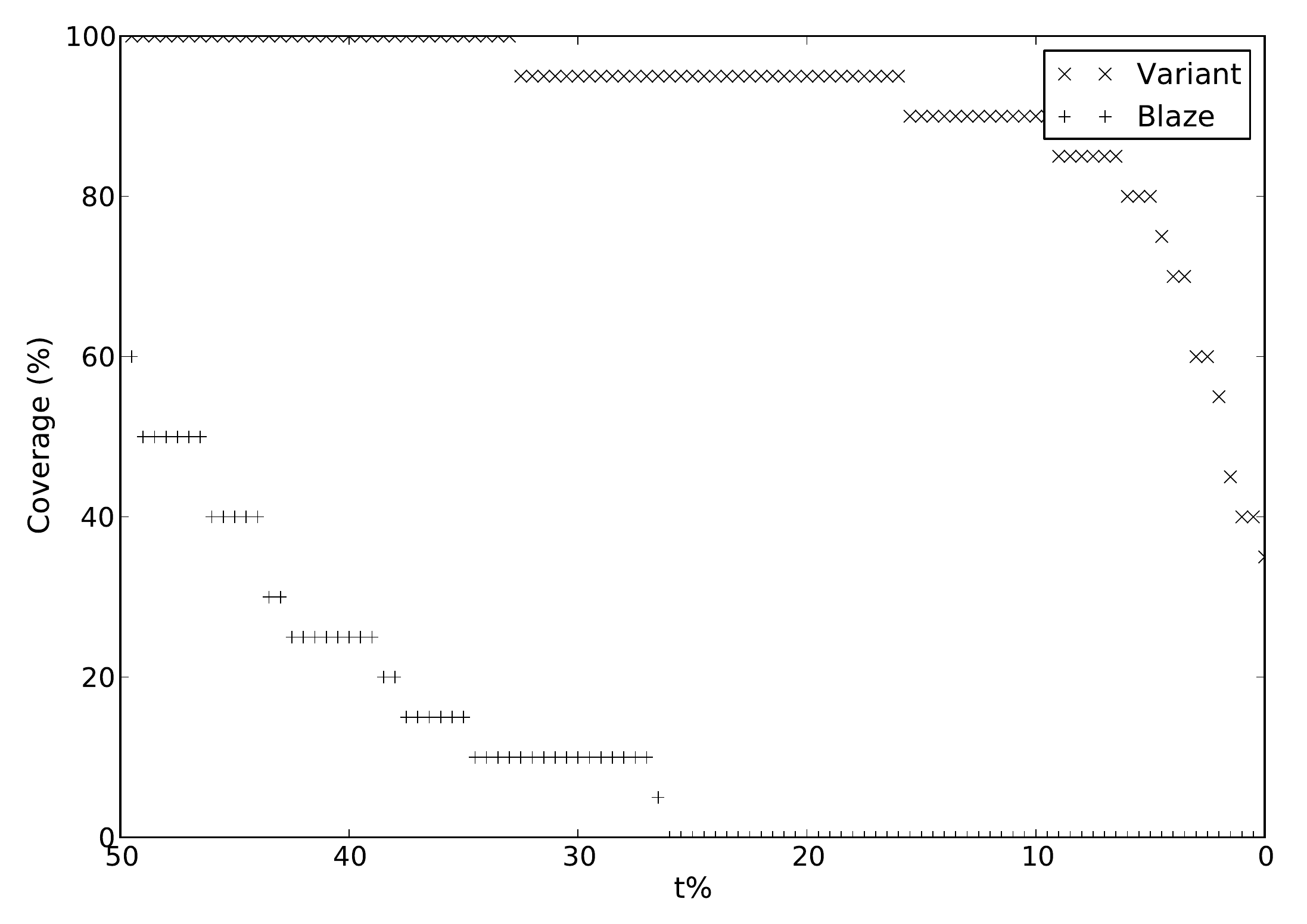}
\end{minipage}%
\begin{minipage}{0.49\linewidth}
\includegraphics[scale=0.32]{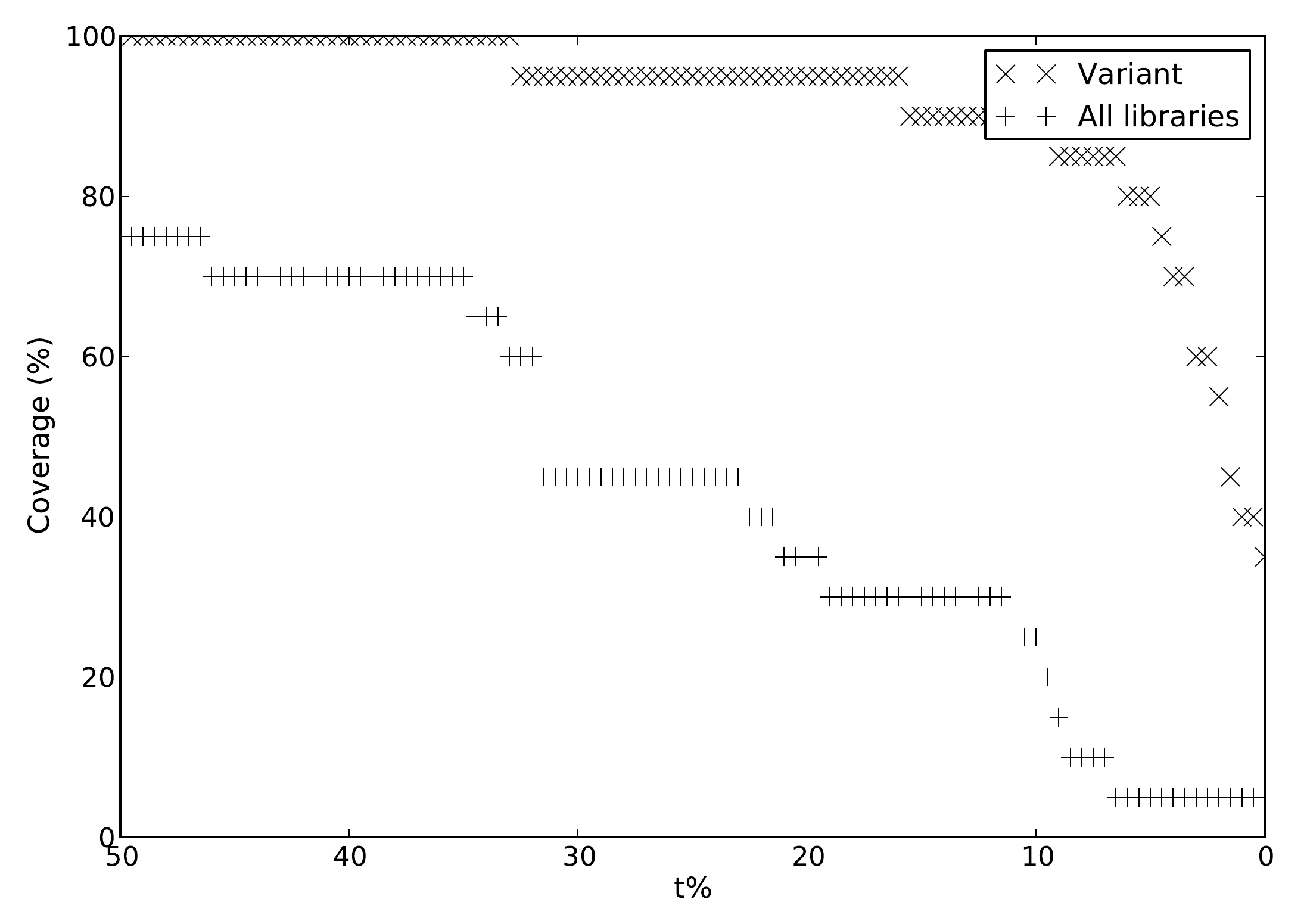}
\end{minipage}%
\caption{The graph shows the coverage obtained for decreasing values of
$t\%$ (closer to the optimum) for the collection of automatically generated
variants and a collection of library routines for the sparse matrix times
vector multiplication kernel on the Xeon 5150 architecture. The graph on the
left only includes routines from the Blaze library, the graph on the right
includes routines from all libraries. As optimum, the optimum found in the
combined collection of variants and library routines is used.}
\label{fig:hpc0-spmatvec}
\end{center}
\end{figure}

The performance of the collection of automatically generated kernels can
also be compared with the performance of collections of routines from sparse
algebra libraries using the coverage metric as defined in the previous
subsection. In this case, the best routine found in the collection of
automatically generated kernels and library routines combined is used as the
optimal routine. The left graph in Figure~\ref{fig:hpc0-spmatvec} shows the
result when this is done for the sparse matrix times vector kernel on the
Xeon 5150 machine and the Blaze library. The x-axis shows the $t\%$ value
that is considered, i.e., is the maximal percentage routines may be distanced
from the optimal implementation.  Smaller values of $t\%$ are closer to the
performance of the optimal implementation.

The coverage in percent can be read from the y-axis, which is the maximal
number of matrices for which a variant is within the selected top group of
implementations. From the figure can be deduced that no Blaze routine is
found that is within 50\% of the optimal implementation for all matrices. In
fact, for $t\%$ values up to 26\%, a routine from the Blaze library never
makes it into the group of best performing variants for any matrix. This
implies not only that the automatically generated kernels always outperform the
Blaze library variants, but also that the automatically generated variants
show a significantly better coverage potential.

The graph on the right in Figure~\ref{fig:hpc0-spmatvec} is a similar graph,
but considers routines from all surveyed sparse algebra libraries instead of
just Blaze. Although the collection of library routines fare better in this
case, still for values of $t$ up to 7\% the performance is fully dominated
by the automatically generated variants. Furthermore, no single library
routine can be found that has a performance within 50\% of the optimal
implementation for all matrices surveyed.

From this graph can also be seen that the automatically generated kernels
persist a 100\% coverage until a $t\%$ value of 34\%, while the library
routines do not achieve a 100\% coverage within 50\% of the optimal
implementation. This indicates that for this particular architecture a
kernel can be found that achieves a reasonable performance for all matrices,
while still significantly outperforming library routines.

This leads us to the following technique to automatically and efficiently
select an all round kernel for a particular architecture.  For a small
random selection of $k$ matrices, all variants are generated and run on the
target architecture. Subsequently, for each of the matrices in the selection
the optimal variant is determined. As a next step, a variant is chosen with
an execution time that is a small, certain percentage $t$ from the optimum
for all matrices in the selection. This kernel has a 100\% coverage for the
selection of matrices and will deliver reasonable performance for any other
matrix.

\begin{table}
\begin{minipage}{0.49\linewidth}
\centering
\begin{tabular}{|l||c|c|}
\hline
(a) & \textbf{Xeon 5150} & \textbf{Xeon E5} \\
\hline
\textbf{SPMV} & 23.7\% & 19.8\% \\
\textbf{SPMM} & 38.1\% & 42.0\% \\
\textbf{TrSv} & 11.2\% & 3.1\% \\
\hline
\end{tabular}
\end{minipage}
\begin{minipage}{0.49\linewidth}
\begin{tabular}{|l||c|c|}
\hline
(b) & \textbf{Xeon 5150} & \textbf{Xeon E5} \\
\hline
\textbf{SPMV} & 9.1\% & 9.5\% \\
\textbf{SPMM} & 25.9\% & 10.0\% \\
\textbf{TrSv} & 13.1\% & 11.0\% \\
\hline
\end{tabular}\end{minipage}
\caption{(a) Minimum average reduction of the execution time of library routines
compared to optimal automatically generated kernel, (b) worst average
reduction of the execution time of automatically selected variant compared
to optimal automatically generated kernel.}
\label{tab:kt-plot}
\end{table}

To evaluate this technique, the average reduction of the execution time of
the kernels found with the above described method and library routines is
compared. The smaller this reduction of the execution time, the better
performance a found kernel has. For each library
routine the average reduction of the execution time has been computed and
the minimum average reduction has been selected and shown in
Table~\ref{tab:kt-plot}(a). As a demonstration of the above described method $k
= 4$ is chosen to obtain a representative selection of matrices and $t\% =
2\%$ to select a top group of variants to consider. This results in a list
of automatically generated kernels that are at most 2\% from the optimal
implementation for all of the $k$ selected matrices. For each of these
kernels, the average reduction of execution time compared to the optimal
kernel has been computed. The worst average found is shown in
Table~\ref{tab:kt-plot}(b), so this is the worst kernel that would be
selected by this method.

In the table can be seen for the sparse matrix times vector multiplication
and sparse matrix times matrix kernels that the worst average reduction of
the execution time of the automatically generated kernels is significantly
smaller than the best average reduction of the execution time of library
routines. For triangular solve, our techniques are on par for the Xeon 5150
architecture and slightly worse for the Xeon E5. As has been described
above, the opportunities for optimization for the triangular solve kernel
are small, leading to the worst average reduction of the execution time of
automatically generated kernels to be larger than that of the library
routine. However, if the average is considered of the kernels selected
through this method, the performance is still on par with that of the
library routine.

By employing this technique a specifically generated kernel can be
selected automatically for a certain architecture by performing run-time
tests with a small amount of matrices.  This generated kernel outperforms
existing library routines for all other matrices.

Note that our approach relies on finding the optimal performance of a kernel
for a given matrix and architecture.  The selection of such an optimal
kernel is an expensive process requiring a vast search space of possible
variants to be explored. However, in the method described above the
optimization is only done once per architecture resulting in a significant
reduction of the optimization overhead while yielding a version of each
kernel which performs substantially better than current approaches, i.e.,
optimized hand-written codes optimized with standard compiler optimizations.

\section{Related Work}
\label{sec:related-work}
Bik and Wijshoff described compiler techniques to automatically generate an
implementation of a computation that operates on sparse matrix structures
from an implementation of that computation expressed in terms of dense
matrices that is supplied to the compiler~\cite{bik-1993,bik-1996}. User
annotations about matrix statistics (e.g. its sparsity) or interactive user
input is used to aid the compiler in selecting an efficient, pre-defined,
sparse storage format for the matrices used in the computation. A major
difference between the work of Bik and Wijshoff and this paper is that in
the case of Bik and Wijshoff a data structure is \emph{selected} from a
pre-defined set of data structures, whereas in this paper we propose
techniques for the automatic \emph{generation} of data structures. The
generation procedure does not depend on any collection of pre-defined
structures or templates.

Mateev et al. proposed a generic programming methodology to bridge the gap
between algorithm implementation API and storage format
API~\cite{mateev-2000}. Algorithms are implemented as generic dense matrix
programs, without considering a particular data storage format. The details
of different, pre-defined, data structure formats are exposed using a
low-level API. Their framework views sparse matrix formats are
indexed-sequential access data structures and uses a restructuring
technology based on relational algebra to convert a high-level algorithm
into a data-centric implementation that exploits characteristics of the
available sparse formats whenever possible. Matrices are considered as
collections of tuples for the purpose of restructuring towards an existing
sparse storage format.

Note that both approaches only restructure towards existing, pre-defined
formats and are not capable of automatically generating a data storage
format like is accomplished by the approach described in this paper.

Marker et al. described a method for the automatic parallelization and
optimization of Dense Linear Algebra for distributed-memory computers called
Design by Transformation (DxT)~\cite{marker-2013}. Their method works by
modeling algorithms in a data-flow graph. The graph contains nodes that
represent redistribution operations or a LAPACK or BLAS function call.
Optimization is carried out by applying graph transformations to find
equivalent graphs that potentially exhibit better performance.

OSKI is a library capable of automatically tuning sparse matrix kernels for
a particular matrix and machine~\cite{vuduc-2005-oski}. The library uses
run-time information to select a data structure and particular
implementation of a kernel that provides improved performance. Contrary to
our approach, OSKI is restricted to a pre-defined set of routines and a
pre-defined set of data structures to tune to. Although this pre-defined set
of routines can be partially generated automatically, the pre-defined set of
data structures is not, as OSKI is not capable of generating new data
structures automatically like is possible with the framework described in
this paper.

Structure splitting is a compiler optimization that transforms an array of
structures into a structure of arrays~\cite{curial-2008}. Whereas this can
be seen as an optimization that modifies how the data is stored, the essence
of the specified data structure remains the same and the way the data is
organized is not substantially modified. Tuple splitting is only one of the
transformation steps in our framework, making the \emph{forelem} framework
far more versatile. Similarly, initial work has been done on run-time
restructuring and linearization~\cite{spek-2008}.  These techniques trace
memory accesses at run-time and subsequently permute the data and/or
transform linked lists to arrays, to attempt to eliminate random memory
accesses. By doing so, the data is permuted to suit the iteration pattern,
or control flow, as has been specified in the program.  Although the order
in which data is stored is addressed, also in this case the way how data is
organized, or grouped, is not changed.  A linked list of entities is
transformed to an array of entities, but the definition of what makes up
this entity is unchanged. So, while these techniques do effectively improve
the performance of certain applications, there is room for much more.

\section{Conclusions and Future Work}
\label{sec:conclusions}
We have introduced a compiler-based framework to enable the automatic
generation of data structures. To do so, a different scheme to specify
programs is proposed, in which the specification of fixed iteration orders
and data structures is avoided. As a result, the compiler can go beyond the
optimization of solely control flow and target the way data is organized and
accessed as well. This is achieved using a collection of transformations,
defined with the \emph{forelem} framework, that target both the loop
structure of the computation as well as the data layout that the code
operates on.

We demonstrated the power of this framework with a case study on the
automatic generation of sparse data structures for sparse matrix routines.
Many different variants of a given (sparse matrix) computation can be
instantiated automatically, where an efficient code routine is combined with
the reassembly of the original sparse matrix data structure into a form that
is better aligned with the computation that is performed.  All data
structures that come forth out of the automatic transformation process
described in this paper are basically the result of the application of a
number of very simple transformations. This includes data structures such as
ITPACK and Jagged Diagonal Storage that have been introduced in the
literature~\cite{kincaid-1986,bai-2000} as smart algorithmic solutions to
optimize sparse matrix times vector computations. Consequently, we can
conclude that these supposedly smart implementations are nothing more than a
succession of very simple basic transformations.

Furthermore, when loop blocking is considered, a multitude of different
hybrid storage schemes can be generated automatically that could impossibly
be pre-defined in a sparse data structure library. Loop blocking is also the
main enabling transformation for the automatic generation of data
distributions for distributed sparse computations. So, with the
transformations described in this paper, it is not only possible to generate
efficient data structures for sparse matrix computations that are executed
on a single processor core, but also to generate efficient data
distributions for distributed sparse computations automatically.

The transformations that enable the generation of sparse data structures to
be automated set up a large transformation search space, leading to much
wider optimization opportunities. This search space can be further extended
with parametric optimizations such as loop unrolling. We have shown that
within this search space, variants of the computation can be found that are
faster than the implementations found in existing sparse algebra libraries.
Additionally, we have shown none of the surveyed existing libraries to
perform consistently better. So, by relying on a single library
implementation performance is never optimal. This further reinforces our
approach to give the compiler control over data structure optimization and
to tailor data structures to input data.

Based on the framework presented in this paper, there are many avenues for
future work. Case studies will be done for other application domains of
these techniques, such as graph data structures, databases and
multi-dimensional storage. This will likely also lead to more generic
transformations to be included in the framework. In particular for
multi-dimensional storage, it is interesting to investigate the
possibilities and benefits of automatically generating hybrid data
structures. Finally, techniques need to be researched to effectively
navigate the search space of possible variants in search of the optimal
routine and to solve the phase ordering problem of the available
transformations.

\bibliographystyle{plain}

\end{document}